\documentclass[]{tCPH2e}

\begin{document}
\doi{10.1080/0010751YYxxxxxxxx}
 \issn{1366-5812}
\issnp{0010-7514}

\jvol{00} \jnum{00} \jyear{2009} \jmonth{November}

\markboth{Soskin, McClintock, Fromhold, Khovanov \& Mannella}{Stochastic webs
and quantum transport}

\articletype{REVIEW}

\title{Stochastic webs and quantum transport in superlattices: \\an introductory review}

\author{S.M.\ Soskin,$^{a,b,c}$$^{\ast}$\thanks{$^\ast$Corresponding author. Email: ssoskin@ictp.it
\vspace{6pt}} P.V.E.\ McClintock,$^c$ T.M.\ Fromhold,$^d$ I.A.\ Khovanov$^e$
and R.\ Mannella$^{f}$\\\vspace{6pt}
$^{a}${\em{Institute of Semiconductor
Physics, National Academy of Sciences of Ukraine, 03028 Kiev,
Ukraine}}\\
$^{b}${\em{Abdus Salam ICTP, 34100 Trieste, Italy}}\\
$^c${\em{Physics
Department, Lancaster University, Lancaster LA1 4YB, UK}}\\
$^d${\em{School of Physics and Astronomy, University of Nottingham,
Nottingham NG7 2RD, UK}}\\
$^e${\em{School of
Engineering, University of Warwick, Coventry CV4 7AL, UK}}\\
$^f${\em{Dipartimento di Fisica, Universit\`{a} di Pisa, 56127 Pisa,
Italy}};
\\\vspace{6pt}\received{v4.0 released 7 November 2009} }

\maketitle

\begin{abstract}
Stochastic webs were discovered, first by Arnold for multi-dimensional
Hamiltonian systems, and later by Chernikov et al.\ for the low-dimensional
case. Generated by weak perturbations, they consist of thread-like regions of
chaotic dynamics in phase space. Their importance is that, in principle, they
enable transport from small energies to high energies. In this
introductory
review, we concentrate on low-dimensional stochastic webs and on their
applications to quantum transport in semiconductor superlattices subject to
electric and magnetic fields. We also describe a recently-suggested
modification of the stochastic web to enhance chaotic transport through it and
we discuss its possible applications to superlattices.
\bigskip

\bigskip
\bigskip

\end{abstract}

\section{Introduction}\label{introduction}

Stochastic webs exist in the phase spaces of Hamiltonian systems, that is, in
the space formed by the coordinates and momenta of a dynamical system evolving
in the absence of dissipation. They consist of a network of very thin
thread-like regions within which the dynamics is chaotic, whereas the dynamics
remains regular everywhere else. Although the concept seems abstract and
mathematical at first sight, stochastic webs are now known to arise in a number
of practical contexts, including for example plasma physics
\cite{Zaslavsky:91}, ultra-cold atoms in optical lattices
\cite{Hensinger:01,Steck:01,Scott:02} and electrons in semiconductor
superlattices (SLs)
\cite{Fromhold:01,Fromhold:04,Fowler:07,Balanov:08,Kuraguchi:02,Greenaway:09,Hyart:09};
they have also been considered in connection with celestial mechanics
\cite{Touma:97}. The importance of the chaotic threads is that they can
transport matter and energy effectively over long distances
\cite{Lichtenberg:89,Afanasiev:90}. In this brief and rather informal review we
aim to introduce the general reader to stochastic webs, explaining what they
are and discussing their recent developments and applications, taking electron
transport in semiconductor SLs as our example.

We start (Section \ref{hamsystems}) from the definition of a Hamiltonian
system, its dimensionality and integrability. Then
(Section \ref{perturbdham})
we consider the effect of
perturbations of the integrable system, which brings us to the concept of a
chaotic (stochastic) layer, in particular related to resonances. The latter
allows us to explain in the beginning of Section \ref{stochwebs} the onset of
the Arnold stochastic web in multi-dimensional systems. The main purpose of
this section is to discuss the more sophisticated nature of low-dimensional
stochastic webs which are, however, still related to the concept of resonance.
In Section \ref{modstochwebs}, we explain the limitations of the transport
though the low-dimensional web and suggest a subtle way of overcoming these
limitations. Finally, in Section \ref{superlattices}, we discuss a rather
unexpected application of the stochastic web concept to quantum electron
transport in nanometer-scale semiconductor SLs in the presence of
electric and magnetic fields. Section \ref{conclusions} draws conclusions.

\subsection{Hamiltonian systems}\label{hamsystems}

Hamiltonian systems play an important role in physics, chemistry, biology and
engineering, and form a fundamental class of dynamical systems
\cite{Arnold:06,Lichtenberg:92,Zaslavsky:07}. They are defined by the following
dynamical equations:

\begin{equation}
\frac {{\rm d}p_i}{{\rm d}t} = - \frac{\partial H}{\partial
q_i},\quad\quad
\frac {{\rm d}q_i}{{\rm d}t} = \frac{\partial H}{\partial p_i}.
\end{equation}

\noindent If the Hamiltonian $H$ does not depend on time $t$, while depending
only on the momenta $\vec{p} \equiv (p_1,\dots,p_N)$ and coordinates
$\vec{q}\equiv (q_1,\dots,q_N)$, then it is called $N$-dimensional. If it also
depends on time $t$, then it has the dimension $N+\frac{1}{2}$. A remarkable
property of any Hamiltonian system is the equality of its full and partial
derivatives with respect to time:

\begin{equation}
\frac {{\rm d}H}{{\rm d}t}=\frac{\partial H}{\partial t}.
\end{equation}

\noindent In particular, for time-independent Hamiltonians,
$H(\vec{p},\vec{q})$ is conserved along the trajectory.

In general, the equations of motion (1) may not be integrable in
quadratures\footnote{When the solution of a differential equation expressible
in terms of a formula involving integrations, it is said to be {\it solvable by
quadrature}.} \cite{Arnold:06,Lichtenberg:92,Zaslavsky:07}, whence the
importance of {\it integrable} systems, i.e.\ those time-independent
Hamiltonian systems for which a transformation
$\{\vec{p},\vec{q}\}\leftrightarrow \{\vec{I},\vec{\theta}\}$ exists such that
\begin{equation}
H(\vec{p},\vec{q})=\tilde{H} (\vec{I}).
\end{equation}
\noindent $I_i$ are called actions while $\theta_i$ are called angles. It
follows from (3) that $\vec{I}$ is conserved:
\begin{equation}
\frac {{\rm d}I_i}{{\rm d}t}=-\frac{\partial \tilde{H} }{\partial \theta_i}=0,
\end{equation}
\noindent while the angles $\theta_i$ change with constant speeds (for a given
$\vec{I}$),
\begin{equation}
\frac {{\rm d}\theta_i}{{\rm d}t}=\frac{\partial \tilde{H}(\vec{I}) }{\partial I_i}\equiv
\omega_i(\vec{I}),
\end{equation}
which
are called frequencies.

Note that angles $\theta_i$ are cyclic variables i.e. $\vec{p}$ and $\vec{q}$
are periodic functions of $\theta_i$ with a period $2\pi$ for any $\theta_i$
\cite{Zaslavsky:91,Arnold:06,Lichtenberg:92,Zaslavsky:07}. Thus, Eqs.\ (4) and
(5) correspond to periodic or quasi-periodic motion. The simplest and most
often used example of an integrable system is a one-dimensional one, which will
be described in more detail below.

\begin{figure}[t]
\center{\includegraphics*[width = 5 cm]{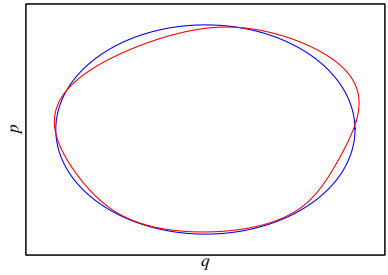}}
\caption {Schematic diagram to show a weak distortion of the majority of
trajectories by a weak time-periodic perturbation: the blue line shows the
trajectory of the unperturbed system,
while the red line shows the stroboscopic Poincar\'{e} section of the trajectory of the perturbed
one.} \label{aba:fig1}
\end{figure}

\subsection{Perturbed Hamiltonian systems}\label{perturbdham}

It is natural to pose the question: what is the effect of a weak perturbation
on an integrable system? For the majority of cases, the answer is given by the
Kolmogorov-Arnold-Moser (KAM) theory \cite{Arnold:06}: most of the trajectories
are just weakly distorted by a weak perturbation while remaining regular. Let
us illustrate this by an example of a one-dimensional system weakly perturbed
time-periodically. In this case, it is convenient to present the trajectory in
the stroboscopic {\it Poincar\'{e} section}
\cite{Zaslavsky:91,Arnold:06,Lichtenberg:92,Zaslavsky:07}, i.e. presenting
states of the system $\{p(t),q(t)\}$ only at the discrete sequence of instants
$t=t_n\equiv t_0+nT$ where $t_0$ is some initial instant, $T$ is the
perturbation period and $n=0,1,2,...$. If the trajectory is \lq\lq just weakly
distorted while remaining regular\rq\rq, then, in particular, the unperturbed
trajectory and the Poincar\'{e} section of the perturbed one have the same
dimension of 1 (i.e. they are just lines), and the same topology, while just
slightly deviating from each other (Fig.\ \ref{aba:fig1}).

\begin{figure}[t]
\begin{center}
\includegraphics*[width = 5 cm]{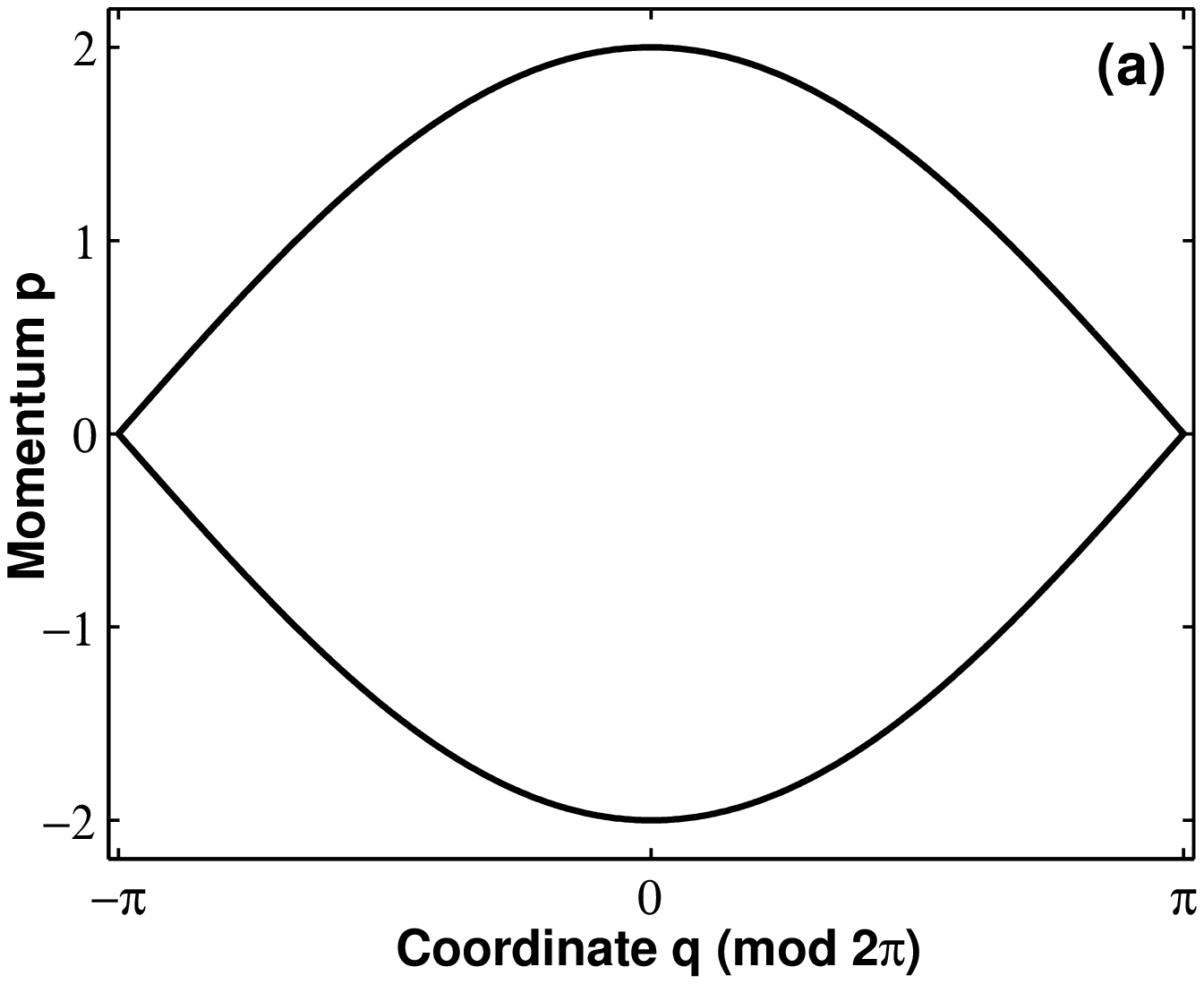} \hspace*{0.5cm}\includegraphics*[width = 5 cm]{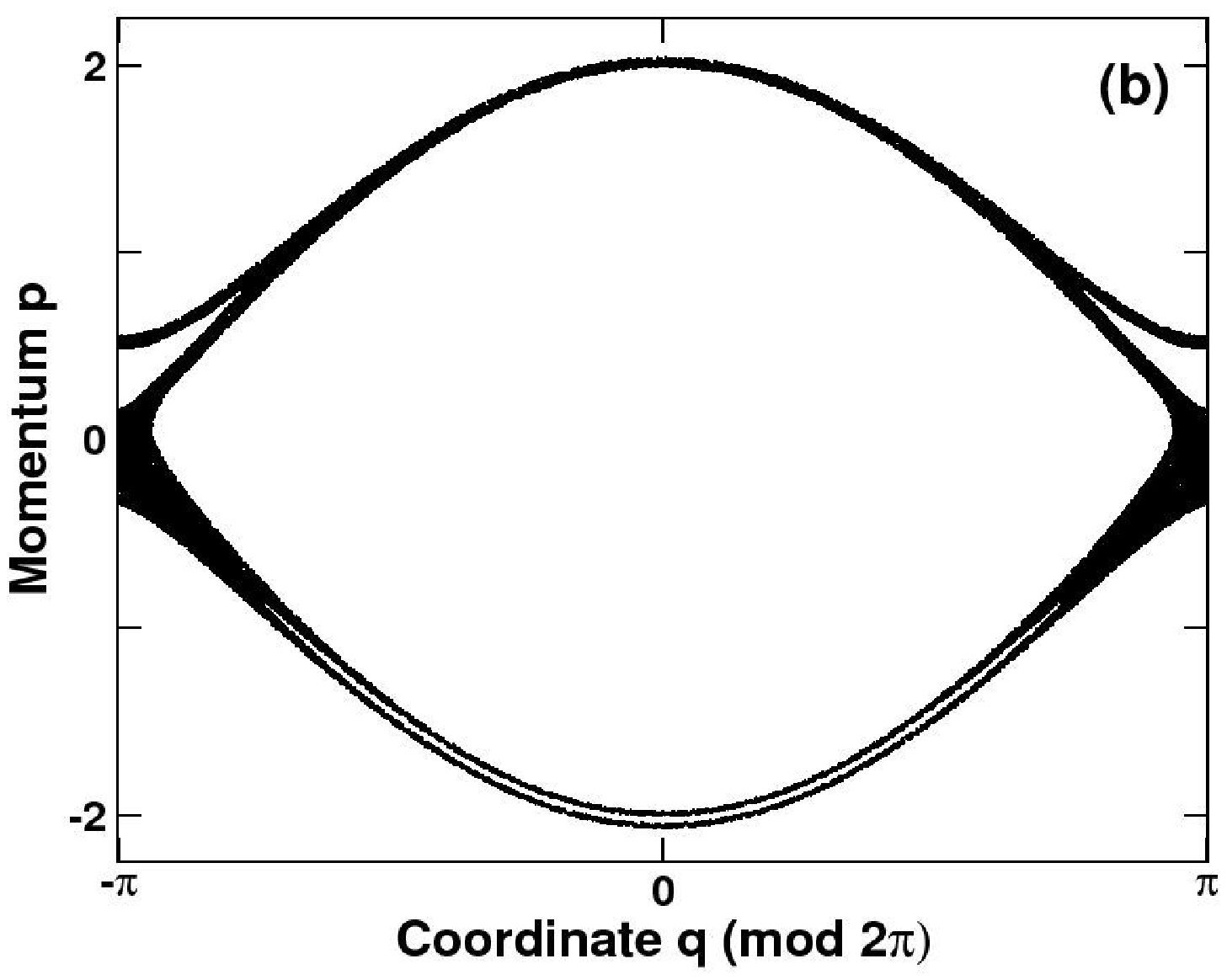}
\caption {(a). The separatrix of the pendulum $H=H_0\equiv p^2/2-\cos(q)$: the
separatrix corresponds to $H=H_s\equiv 1$. (b). The chaotic layer (replacing the
separatrix) in Poincar\'{e} section of the ac-driven pendulum: $H=H_0-0.01q\cos(t)$. } \label{abc:fig2}
\end{center}
\end{figure}

There are, however, two kinds of situation for which KAM-theory is not valid.
The first of these relates to the {\it separatrices} of the unperturbed
systems. Let an integrable system possess a separatrix i.e.\ the line (or
surface, or hyper-surface in the general multi-dimensional case) that separates
trajectories of a different topology in the phase space\footnote{More
rigorously, the separatrices may be defined as follows \cite{Gelfreich:01}. Let
the integrable system possess a saddle i.e.\ a hyperbolic point in the
one-dimensional case (i.e. an unstable stationary point with an exponential
dynamics of trajectories approaching it), or a hyperbolic invariant torus in
higher-dimensional cases. The stable (incoming) and unstable (outgoing)
manifolds are called {\it separatrices}.}: e.g.\ in the example shown in Fig.\
\ref{abc:fig2}(a), the separatrix separates closed trajectories (corresponding
to oscillations inside the separatrix loops) from open trajectories
(corresponding to the running coordinate below or above the separatrix loops).
If the system is perturbed time-periodically\footnote{In multi-dimensional
cases, a time-independent perturbation also may give rise to the invalidity of
the KAM-theory near the separatrix.}, then the separatrix is replaced by a
chaotic trajectory. In Poincar\'{e} section, the chaotic trajectory lies within
a {\it chaotic layer} (Fig.\ \ref{abc:fig2}(b)): the latter has a complicated
(fractal) structure but its outer boundaries are well defined and the region
delineated by these boundaries has the dimension 2, unlike the dimension 1 of
regular trajectories. Thus, even the appearance of the Poincar\'{e} section
allows us to distinguish immediately between regular and chaotic trajectories,
unless of course the width of the chaotic layer is less than the accuracy
provided by the numerical integration of equations of motion. The theoretical
prediction of the width in energy of the chaotic layer has a long and rich
history. Its description on a physics level of rigour may be found in the book
by Zaslavsky \cite{Zaslavsky:07}. Studies that are more mathematically rigorous
have recently been reviewed \cite{Piftankin:07}. The maximum width of the layer
and other significant features (high peaks) of the width as function of the
perturbation frequency have recently been described
\cite{Soskin:09a,Soskin:09b}.

Another characteristic situation where the KAM-theory is invalid relates to
{\it resonances}, i.e.\ to areas of the phase space where at least one of the
following conditions holds

\begin{eqnarray}
&&
n\omega_i(\vec{I}_r)=
m\omega_j(\vec{I}_r),
\\
&&
n,m=\pm 1,\pm 2, \pm 3, \dots,
\nonumber
\\
&&
i,j=1,2,3, \dots,N, \quad\quad i\neq j,
\nonumber
\\
&&
N=2,3,4, \dots,
\nonumber
\end{eqnarray}

\noindent or

\begin{eqnarray}
&&
n\omega_i(\vec{I}_r)=
l\omega_f,
\\
&&
n=\pm 1,\pm 2, \pm 3, \dots,
\nonumber
\\
&&
i=1,2,3, \dots,N-\frac{1}{2},
\nonumber
\\
&&
N=\frac{3}{2},\frac{5}{2},\frac{7}{2},\dots,
\nonumber
\end{eqnarray}

\noindent where $\omega_f$ is the frequency of the corresponding time-periodic
perturbation while $l$ is the number of the Fourier harmonic that may exist for
the time-periodic perturbation (e.g.\ for a monochromatic perturbation, only
$l=1$ is relevant).

The rigorous treatment of motion in the resonance range is rather complicated,
being related to the Poincar\'{e}-Birkhoff theorem and homoclinic and
heteroclinic trajectories and tangencies \cite{Haller:99,Tabor:87}. We do not
consider it here. Rather, we give a brief interpretation of the
resonance-related chaos in physical terms\footnote{This was given for the first
time by Chirikov \cite{Chirikov:59} (a clear presentation of the issues in
question can be found e.g.\ in \cite{Zaslavsky:07}).}. For the sake of clarity,
consider an ac-driven 1D Hamiltonian system whose frequency of eigenoscillation
$\omega$ increases monotonically with the energy of eigenoscillation
$E\equiv H_0(p,q)$, while
the perturbation frequency $\omega_f$ exceeds the minimum of $\omega(E)$, as
shown in Fig.\ \ref{abd:fig3}:

\begin{figure}[t]
\center{\includegraphics*[width = 5.3 cm]{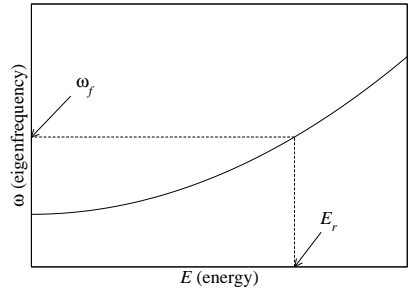}}
\caption {A schematic diagram showing the dependence of eigenfrequency $\omega$
on the energy $E$ of an eigenoscillation, and the meaning of the resonance
energy $E_r$.} \label{abd:fig3}
\end{figure}

\begin{eqnarray}
&&
H=H_0(p,q)-hq\cos(\omega_ft),
\\
&&
h\ll 1,
\nonumber
\\
&&
\omega_f>\omega_0\equiv\min\{\omega(E)\}.
\nonumber
\end{eqnarray}

\noindent Then there necessarily exists an energy $E_r$ such that the resonance
condition (7) with $n=l=1$ holds true:
\begin{equation}
\omega(E_r)=\omega_f.
\end{equation}

\noindent Consider motion for energies close to $E_r$. Let us transform from
variables $\{p,q\}$ to action-angle variables $\{I,\theta\}$, so that the
Hamiltonian becomes

\begin{eqnarray}
H(p,q) &\equiv& \tilde{H}(I,\theta)= \int_{I_{\min}}^{I}{\rm d}\tilde{I}\omega(\tilde{I})
- h\sum_nq_n(I)\cos(n\theta)\cos(\omega_ft)
\nonumber
\\
&\equiv&
\int_{I_{\min}}^{I}{\rm d}\tilde{I}\omega(\tilde{I})-
\frac{1}{2}hq_1\cos (\theta-\omega_ft)+ . . .
\nonumber
\\
&\equiv&
\tilde{H}_0(I,\tilde{\theta}\equiv \theta-\omega_ft)+V_f(I,\tilde{\theta},t),
\\
I(p,q)&\equiv& I(E)=\frac{1}{2\pi}\oint p(q,E) \;{\rm d} q,
\quad\quad
E\equiv H_0(p,q),
\nonumber
\\
\theta(p,q)&=& \omega(E)\int^q \frac{1}{p(\tilde{q},E)} \;{\rm d}
\tilde{q},
\nonumber
\\
q_n &\equiv& q_n(I)=\frac{1}{2\pi}\int_{0}^{2\pi}{\rm d}\theta q
\cos(n\theta).
\nonumber
\end{eqnarray}

\noindent Here, the dots \lq\lq $\dots$\rq\rq denote terms that vary with time
much faster than the preceding term: they are denoted in the next equality as
$V_f$. Thus, allowing for the resonance condition (9), we may introduce the
slow angle $ \tilde{\theta}\equiv \theta-\omega_ft $ and present the original
Hamiltonian as a sum of an \lq\lq autonomous\rq\rq part
$\tilde{H}_0(I,\tilde{\theta})$ and the time-dependent (fast-oscillating) part $V_f(I,
\tilde{\theta},t)$. It is easy to check by direct substitution into the
corresponding Hamiltonian equations of motion that the dynamics of the
variables $I$ and $\tilde{\theta}$ is governed by the Hamiltonian
\begin{equation}
\tilde{ \tilde{H}}= \tilde{H}-\omega_fI.
\end{equation}
\noindent Taking into account that the perturbation is small ($h\ll 1$) and,
therefore, that the variation of $I$ around the resonance value $I_r\equiv
I(E_r)$ is also small, we may approximate the function $\omega(I)$ near the
resonance value as
\begin{eqnarray}
&&
\omega(I)\approx\omega_f+ \omega_r^{\prime}(I-I_r),
\\
&&
\omega_r^{\prime}\equiv \frac{{\rm d} \omega}{{\rm d} I}\left |_{I=I_r}\right ..
\nonumber
\end{eqnarray}
From (10)-(12), we ultimately obtain the approximate auxiliary Hamiltonian governing the
dynamics of $\{I,\tilde{\theta} \}$:
\begin{eqnarray}
\tilde{\tilde{H}} ( I,\tilde{\theta},t)&=& \frac{1}{2}\omega_r^{\prime}(I-I_r)^2-\frac{1}{2}hq_1\cos(\tilde{\theta})+V_f( I,\tilde{\theta},t)
\nonumber
\\
&\equiv& \tilde{\tilde{H}}_0 ( \tilde{I}\equiv I-I_r,\tilde{\theta})+V_f.
\end{eqnarray}

\noindent It represents the sum of a pendulum-like\footnote{$\tilde{I}$ plays
the role of a generalized velocity while $\tilde{\theta}$ plays the role of the
generalized coordinate. Note that the generalized potential contains the small
multiplier $h$, so that the maximal absolute value of the \lq\lq velocity\rq\rq
$\tilde{I}$ is proportional to $\sqrt{h}$ and, therefore, is small too.}
autonomous Hamiltonian $\tilde{\tilde{H}}_0 ( \tilde{I},\tilde{\theta})$ and a
time-periodic (fast-oscillating) part $V_f$ that plays the role of a perturbation. The
pendulum-like part $\tilde{\tilde{H}}_0 $ possesses a separatrix (cf.\ Fig.\
\ref{abc:fig2}(a)) while the perturbation-like part $V_f$ tends to destroy the
separatrix, replacing it with an exponentially narrow chaotic layer.

Thus we have shown that a resonance is necessarily associated with a narrow
chaotic layer.

\section {Stochastic webs}\label{stochwebs}

In the example considered above, the chaotic layer associated with the
resonance provides only a narrow ($\propto\sqrt{h}$) variation of energy (or,
equivalently, of action). Thus, there is no significant transport in energy.
Let us pose a question: {\it could there be situations when a perturbation
provides for chaotic transport through a large range of energies?} We now
describe the three stages of conceptual evolution that led to a positive answer
to this question.

\subsection {Multi-dimensional web}

First, Arnold showed in 1964 \cite{Arnold:64} through rather simple topological
arguments (also presented clearly in \cite{Zaslavsky:07}) that, if the system
is multi-dimensional (namely, if $N\geq 5/2$), and if the so called
non-degeneracy condition $\det(\partial^2H/\partial I_i\partial I_j)\neq 0$ is
fulfilled (in other words, if the system is sufficiently nonlinear), then
resonances necessarily intersect with each other, forming an infinite {\it web}
in the phase space along which an exponentially slow chaotic diffusion occurs.

\subsection {Low-dimensional webs}

Secondly, Chernikov et al.\ published an important series of papers in the late
1980s. We shall review just three of the more important of them, concentrating
on the model of a harmonic oscillator subject to a plane wave, which will be
relevant to our discussion of semiconductor SLs below. A good review
of a broad range of the early work on low-dimensional stochastic webs may be
found in \cite{Zaslavsky:91}; more recent work is reviewed in
\cite{Zaslavsky:07b} (see also \cite{Zaslavsky:07}).

The main idea of Chernikov et al.\ is that a stochastic web may arise even in
low-dimensional systems ($N=\frac{3}{2};2$) provided that the non-degeneracy
condition is lifted, in other words, in this case, if
\begin{equation}
\frac{{\rm d} \omega}{{\rm d} I}=0,
\end{equation}
\noindent while the perturbation in the equation of motion is resonant and coordinate-dependent.

\subsubsection {Cobweb}

\begin{figure}[t]
\center{\includegraphics*[width = 6. cm]{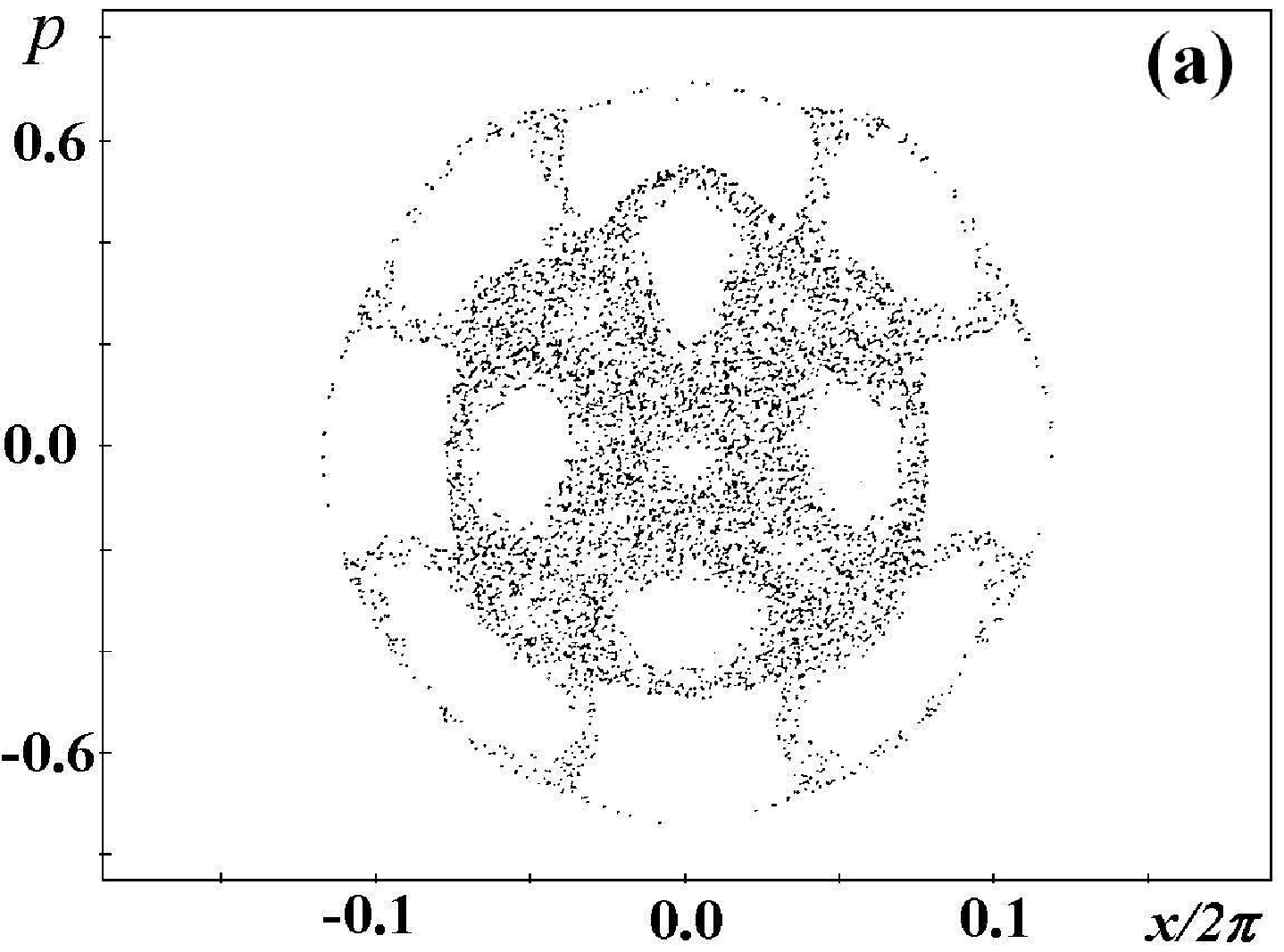} \hspace{1cm}\includegraphics*[width = 6. cm]{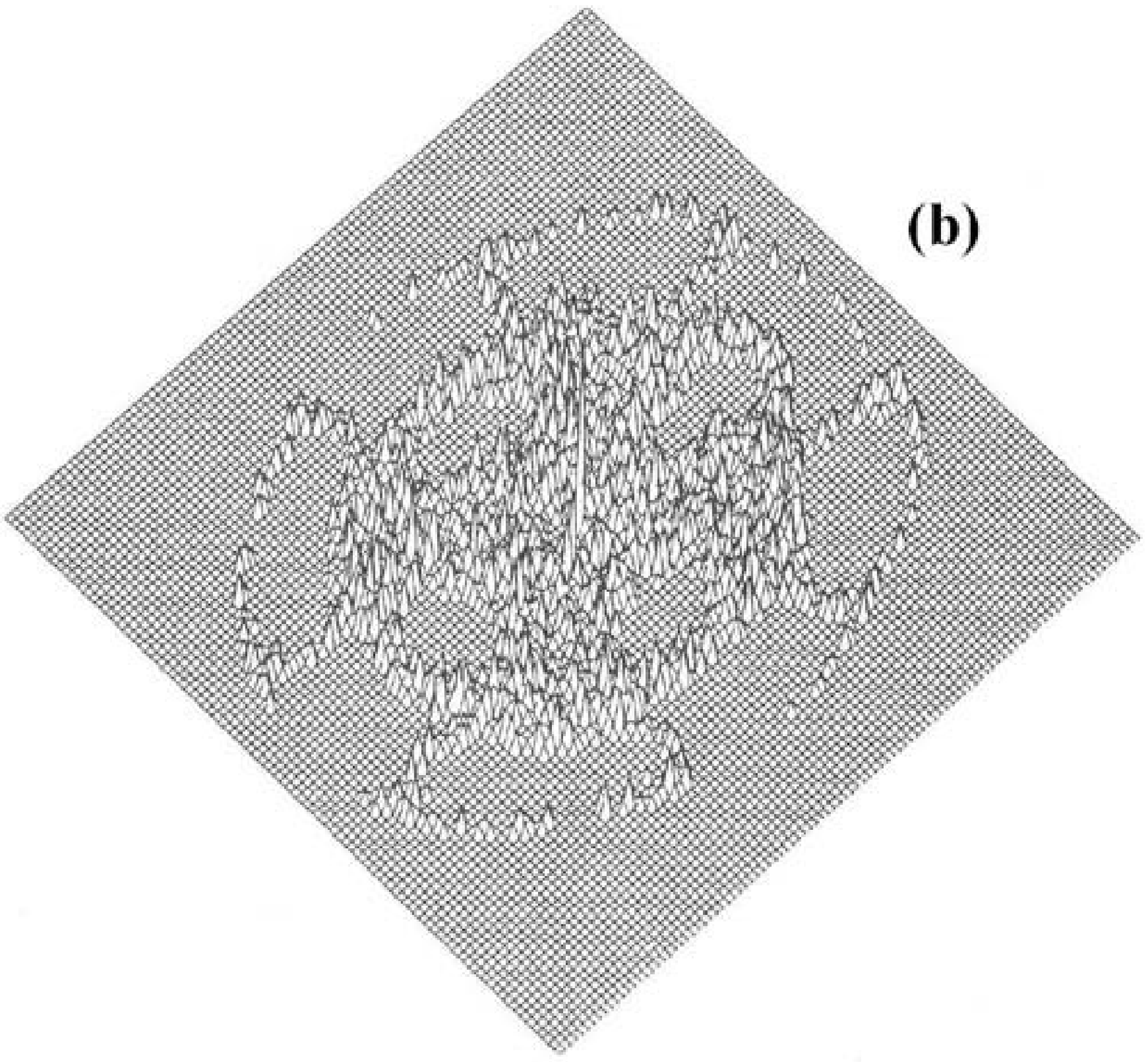}}
\caption {Stochastic web for the system (15), as obtained numerically with $\omega_0=1$, $\nu=4$,
$k=15$, $\epsilon\omega_0^2/k=0.1$ \cite{Chernikov:87a}. The integration time
is $2\times 10^4$ periods of oscillation $2\pi/\omega_0$. (a). Poincar\'{e}
section. (b). The corresponding probability distribution.} \label{abc:fig5}
\end{figure}

\begin{figure}[tb]
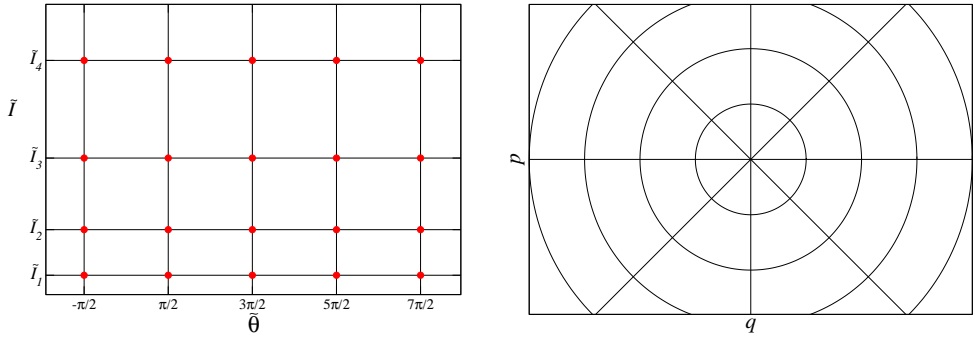

\center{\includegraphics*[height = 4.4cm]{Fig5l.eps}\hspace{0.5cm}
\includegraphics*[height = 4.4 cm]{Fig5r.eps}}
\caption {Left panel: schematic representation of the grid-like separatrix of the Hamiltonian
$\tilde{H}_s$, as defined in Eqs. (20); saddles are shown by dots.
Right panel: schematic representation of the same separatrix in Poincar\'{e}
section $p-q$.} \label{abe:fig6}
\end{figure}

We now consider the best known example of a low-dimensional stochastic web, the
skeleton of which in $p-q$ Poincar\'{e} section has a form resembling that of a
cobweb (Figs.\ \ref{abc:fig5}(a), \ref{abe:fig6}(b)).
We suppose that a harmonic oscillator is perturbed\footnote{For parameters, we
use the same notation as Zaslavsky \cite {Zaslavsky:91} and Chernikov et al.\
\cite{Chernikov:87a} while the coordinate is denoted as $q$ (instead of $x$ in
\cite {Zaslavsky:91,Chernikov:87a}: cf. Fig. 4(a)) in order to match the
notation in other sections and in some figures from other works reproduced
below.} by a resonant plane wave \cite{Chernikov:87a}:
\begin{eqnarray}
&&
\ddot{q} + \omega_0^2q=\epsilon\frac{\omega_0^2}{k}\sin(kq-\nu t),
\\
&&
\nu=n\omega_0,\quad n=1,2,3, . . .
\nonumber
\end{eqnarray}
\noindent This particular model has a number of applications, especially to
plasma physics \cite{Zaslavsky:91} and to semiconductor SLs, as shown
below in Sec.\ \ref{superlattices}. In order to understand the origin of the
stochastic web shown in
Fig.\ \ref{abc:fig5},
we


\begin{itemize}

\item[(i)]
transform to polar coordinates $\{\rho,\theta\}$ or,
    equivalently, to action-angle variables $\{I,\theta\}$:
\begin{eqnarray}
&&
q=\rho \sin(\theta),\quad p\equiv \dot{q}=\omega_0\rho\cos(\theta),
\\
&&
\rho\equiv\sqrt{\frac{2I}{\omega_0}},
\nonumber
\end{eqnarray}

\item[(ii)]
make use of the formula \cite{Abramowitz:70}
\begin{equation}
\cos(x\sin(\theta)-y) = \sum_{m=0}^{\infty} J_m(x)\cos(m\theta - y)
\end{equation}

\noindent where $J_m (x) $ is a Bessel function of the $m$th
order\footnote{Note that $J_m (x) $ is an oscillatory function of $x$ with
gradually decreasing amplitude as $x$ increases. At $x\sim 1$, the period
of oscillation is $\sim 2\pi$ while the amplitude is $\sim 1$. For large
$x$, the Bessel function asymptotically approaches the function
$\sqrt{2/(\pi x) }\cos(x-(2m+1)\pi/4) $. }.

\end{itemize}

\noindent
Using (16) and (17), it is not difficult to show that the Hamiltonian
of a harmonic oscillator perturbed by a plane wave can be represented in
action-angle variables as
\begin{equation}
H(I,\theta,t) = \omega_0I + \epsilon\frac{\omega_0^2}{k^2}\sum_m J_m(k\rho(I))\cos(m\theta - \nu t).
\end{equation}
Note that, due to the resonance condition $\nu=n\omega_0$ in (15), the term in
the sum in (18) corresponding to $m=n$ is nearly constant compared to other
terms in the sum. So, similarly to the case of nonlinear resonance considered
in Section \ref{perturbdham} above, it is this term that provides the major
contribution to the dynamics. The other terms in the sum play the role of
fast-oscillating perturbations. So we again introduce a slow variable, the
angle $\tilde{\theta}\equiv n\theta-\nu t$. It is also convenient to introduce
the normalized action $\tilde{I}\equiv I/n$. The dynamics of the slow variables
$\{ \tilde{I},\tilde{\theta}\}$ is then governed by the auxiliary Hamiltonian
$\tilde{H}\equiv nH-\nu \tilde{I}$ (as may readily be checked by direct
substitution into the Hamiltonian equations of motion). Hence

\begin{eqnarray}
\dot {\tilde I} &=& -\frac{\partial \tilde H}{\partial \tilde \theta}, \quad \dot {\tilde \theta} = \frac{\partial \tilde H}{\partial \tilde I},
\\
\tilde{\theta}&\equiv& n\theta-\nu t, \quad
\tilde{I}\equiv I/n,
\nonumber
\\
\tilde H &=& \tilde H_s + \tilde V_f,
\nonumber
\\
\tilde H_s &\equiv& \tilde H_s(\tilde I, \tilde \theta) = \frac{\epsilon n}{k^2}\omega_0^2 J_n(k\rho(\tilde I))\cos(\tilde \theta),
\nonumber
\\
\tilde V_f &\equiv& \tilde V_f(\tilde I, \tilde \theta, t) =
\quad\frac{\epsilon n}{k^2}\omega_0^2 \sum_{m\neq n}J_m(k\rho(I))\cos\left(\frac{m}{n}\tilde \theta-\left(1-\frac{m}{n}\right)\nu t\right).
\nonumber
\end{eqnarray}
Thus, $\tilde{H}_s$ is an autonomous Hamiltonian that determines the main
features of the motion of $\{ \tilde{I},\tilde{\theta}\}$, while $\tilde{V}_f$
plays the role of a fast-oscillating perturbation.

It is straightforward to show that the autonomous Hamiltonian $\tilde{H}_s$
possesses a single infinite grid-like separatrix corresponding to the zero
value of $\tilde{H}_s$ (Fig.\ \ref{abe:fig6}, left panel). The vertical filaments of the
grid correspond to $\tilde{\theta}$ being equal to odd multiples of $\pi/2$
while the horizontal filaments correspond to zeros of the relevant Bessel
function,
\begin{eqnarray}
&&
{\rm separatrix:} \quad \tilde{H}_s=0:
\\
&&
\tilde{\theta}=(2j+1)\frac{\pi}{2},\quad j=0,\pm 1, \pm 2, \dots,
\nonumber
\\
&&
\tilde{I}=\tilde{I}_i,
\quad i=0,1,2, \dots,
\nonumber
\\
&&
J_n(k\rho(\tilde{I}_i))=0.
\nonumber
\end{eqnarray}

\noindent Note that the grid-like separatrix does not depend on the amplitude
of the original perturbation. Rather its form is an inherent property of the
harmonic oscillator driven by the resonant plane wave.

The fast-oscillating term $\tilde{V}_f$ replaces this grid-like separatrix by
the narrow chaotic layer. If the separatrix (20) is represented in the
Poincar\'{e} section $p-q$, it takes precisely the cobweb form shown
schematically in the right-hand panel of Fig.\ \ref{abe:fig6}. Thus we have
achieved the primary goal of this subsection, to explain the onset of the
cobweb-like stochastic web.

\subsubsection{Width of the cobweb}

\begin{figure}[tb]
\center{\includegraphics*[width = 6 cm]{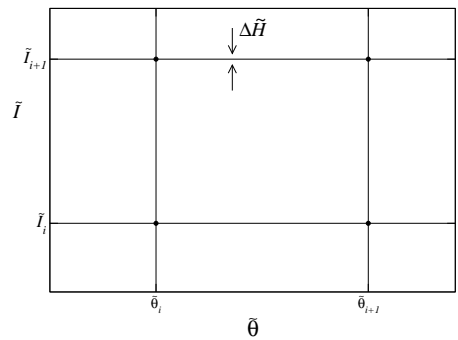}}
\caption {Schematic diagram showing the width of the chaotic layer of the web.}
\label{abe:fig7}
\end{figure}

Transport through the web is affected, not only by the shape of the web's
skeleton, but also by its width, i.e.\ the width of the chaotic layer (Fig.\
\ref{abe:fig7}). An exact calculation of the width is a complicated task that
we will not undertake here. Rather, we will make a rough estimate sufficient to
lead us to definite qualitative conclusions.

\begin{figure}[tb]
\center{\includegraphics*[width = 6 cm]{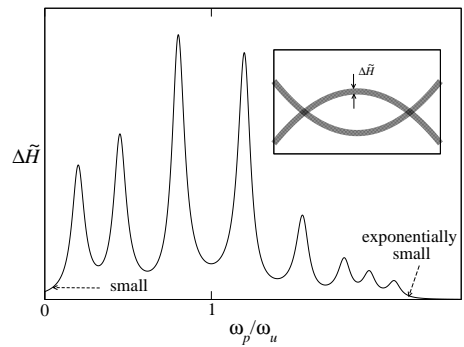}}
\caption {Typical dependence (schematic) of the width of a separatrix chaotic
layer on the ratio between the frequency of perturbation and the frequency of
the small-amplitude oscillation of the unperturbed system.} \label{abe:fig8}
\end{figure}

Before doing so, we make a general comment about the width in the case of a
1D system with a separatrix that is being perturbed by a time-periodic
perturbation. The width depends strongly on the ratio $\omega_f$ between the
frequency of the perturbation $\omega_{\rm perturbation}$ and the frequency of
small-amplitude eigenoscillations $\omega_{\rm unperturbed}$. A schematic
representation of the typical dependence is shown in Fig.\ \ref{abe:fig8}. This
figure will be discussed in more detail in Section \ref{modstochwebs}. In the
present context it is sufficient to emphasize that, if $\omega_f$ is large,
then the width of the chaotic layer is {\it exponentially narrow}.

Let us now turn to the case of the web. As seen from (19), the characteristic
frequency of the perturbation $\tilde{V}_f$ is $\sim \omega_0$. On the other
hand, the unperturbed Hamiltonian $\tilde{H}_s$ is proportional to the small
parameter $\epsilon$. Therefore, even without a careful analysis of its
oscillations, it is clear that the frequency of oscillation in any cell of its
grid-like separatrix is also small. Thus, we conclude that the ratio $\omega_f$
between the perturbation frequency and the eigenfrequency is large, so that the
width of the layer should be exponentially small. This conclusion is confirmed
both by careful theoretical analysis and by numerical simulations
\cite{Zaslavsky:91,Chernikov:87a}.

Moreover, the analysis of oscillations near the elliptic points inside the
cells of the separatrix of $\tilde{H}_s$ shows that, for cells far from the
centre, the frequency of oscillation possesses the following
property\cite{Zaslavsky:91,Chernikov:87a}
\begin{equation}
\omega_{\rm unperturbed}\propto\frac{\epsilon}{I^{3/4}},
\end{equation}
i.e.\ it decreases as the distance from the centre increases. Conversely, the
ratio $\omega_f$ increases. This means that the width of the layer decreases
exponentially quickly as the distance from the centre of the web increases.
This conclusion is confirmed by Fig.\ \ref{abc:fig5} above: even for the
moderate $\epsilon$ used in this case, the width of the layer markedly
decreases as the distance from the centre grows.

\subsubsection{Inexact resonance}

\begin{figure}[tb]
\center{\includegraphics*[width = 7.5 cm]{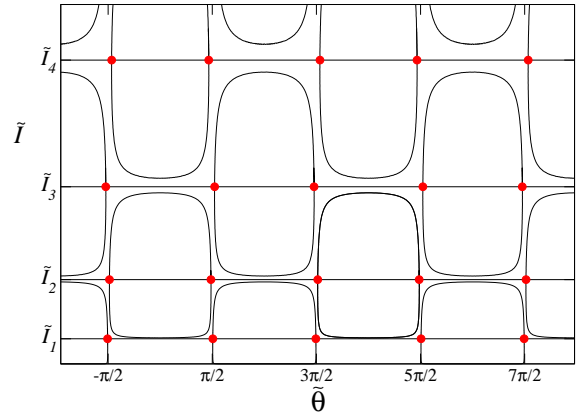}}
\caption {The separatrix for inexact resonance: the single grid-like separatrix
of Fig.\ \ref{abe:fig6} is replaced by a set of separatrices that are
distinctly separated from each other. The parameters used to compute the
separatrix were: $\Delta \omega = -0.001 \neq 0$ (see Eq. (22)), $\epsilon = 0.573$
and $n=k=1$.} \label{abe:fig9}
\end{figure}

Natural questions to ask in relation to the cobweb are: what happens if the
oscillator differs slightly from an ideal harmonic oscillator;  and what
happens if the resonance is inexact? The answers were given by Chernikov et
al.\ \cite{Chernikov:88} (see also \cite{Zaslavsky:91}). They found that the
effects of anharmonicity and inexact resonance are in fact similar. So in what
follows we shall, for the sake of brevity, consider only the inexactness of the
resonance:
\begin{equation}
\nu=n\omega_0+\Delta \omega, \quad \Delta \omega\ll \omega_0.
\end{equation}
\noindent In this case, the autonomous resonance Hamiltonian reads as (cf.\
(19))
\begin{equation}
\tilde H_s = \Delta \omega \tilde I + \frac{\epsilon n}{k^2}\omega_0^2 J_n(k\rho(\tilde I))\cos(\tilde \theta)
\end{equation}
\noindent As before, there are saddle points corresponding to different values
of $\tilde{I}$, namely different roots of the equation
$J_n(k\rho(\tilde{I}))=0$. But this means that, unlike the resonance case
($\Delta \omega=0$), the values of $\tilde{H}_s$ at the saddles corresponding
to different $\tilde{I}$ are themselves different. This means that the single grid-like
separatrix splits into infinitely many different separatrices (Fig.\ \ref{abe:fig9}).

In order for at least two lowest separatrices to be connected, allowing chaotic
transport within a unified structure (a web of a finite size, in the $p-q$
plane), the width of the chaotic layer should be more than, or of the order of,
the difference in $\tilde{H}_s$ between the two lowest separatrices:
\begin{equation}
\Delta\tilde H \stackrel{\sim}{>}|\tilde H_s(\tilde {I}_2)-
\tilde H_s(\tilde {I}_1)|\sim 2\pi|\Delta \omega|.
\end{equation}
\noindent Because $\Delta\tilde H $ is exponentially narrow, as discussed in
Section \ref{stochwebs}.2.2 above, the inequality (24) means that the
stochastic web may be formed only if the perturbation frequency lies in an {\it
exponentially small vicinity} of the resonance.

\subsubsection {Uniform web}

\begin{figure}[tb]
\center{\includegraphics*[width = 4 cm]{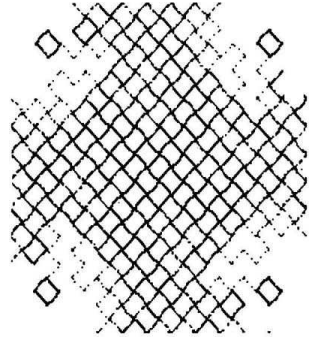}}
\caption {Example of a uniform web in $p-q$ Poincar\'{e} section
\cite{Zaslavsky:91}. } \label{abe:fig10}
\end{figure}

As already demonstrated above, the cobweb cannot in practice provide transport
to arbitrarily large energies because of the exponentially fast decrease in
the width of the chaotic layer with distance from the centre of the web. This
limitation is overcome in another type of the stochastic web, called the {\it
uniform web} \cite{Chernikov:87b} (see also \cite {Zaslavsky:91}). Here,
instead of being perturbed by a plane wave, the harmonic oscillator is
perturbed by short kicks that are periodic in space and time such that the kick
frequency is equal to the eigenfrequency of the oscillator or to one of its
multiples:
\begin{eqnarray}
&&\ddot{q}+q=-\epsilon\sin(kq)\sum_{n=-\infty}^{\infty}\delta(t-nT),
\\
&&
T=\frac{2\pi}{\nu}, \quad \nu=1,2,3,\dots
\nonumber
\end{eqnarray}
The web then covers the whole phase space uniformly, as shown in Fig.\
\ref{abe:fig10}.

We note however that the width of the chaotic layer is still exponentially
small if the amplitude of the perturbation is small
\cite{Zaslavsky:91,Chernikov:88}.

\section {Modified stochastic webs}\label{modstochwebs}

It is clear from the above discussion that a serious limitation affecting
transport through any chaotic web is the exponential narrowness of the web's
chaotic layer, which leads to exponentially slow transport. Soskin et al.\
\cite{Soskin:09b,Soskin:09c,Soskin:09d} recently suggested a way of overcoming
this problem by making a subtle modification of the webs leading, in turn, to
exponential growth in the width of the chaotic layer. We shall demonstrate this
idea on our example of the cobweb, both because it is relevant to the
application to the semiconductor SLs and because, in this case, it
also leads to a dramatic increase in the size of the web.

\subsection{Exact resonance case}

\begin{figure}[tb]
\begin{center}
\includegraphics*[width = 8.2 cm]{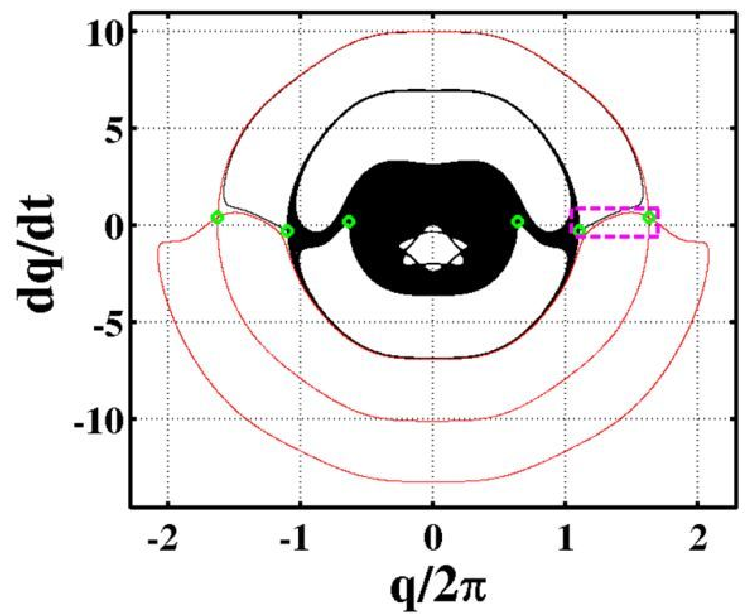} \hspace{0.3cm}
\includegraphics*[width = 7.3 cm]{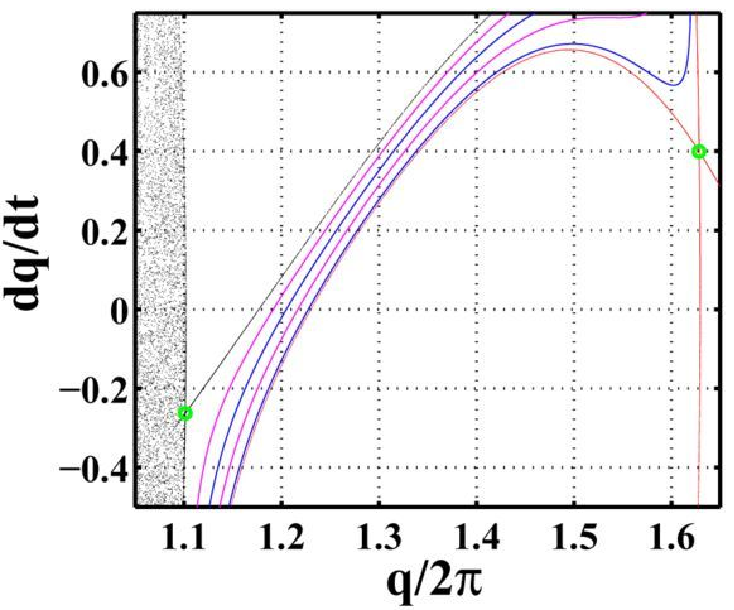}
\caption {Left figure. Poincar\'{e} section for the system $\ddot{q}
+q=\cos(q-t)$. The saddles (marked by green circles) have been found
numerically. Four inner saddles belong to one and the same chaotic trajectory
(shown in black) which forms two-and-a-half inner loops of the stochastic web.
Two remaining (outer) saddles generate another (shown in red) chaotic
trajectory which covers a very thin chaotic layer and is distinctly separated
from the black chaotic trajectory. Right figure. The area within the dashed
magenta rectangle of the left figure is shown on a larger scale. Apart from the
black and red chaotic trajectories, we show (in magenta and blue) examples of
regular trajectories (corresponding to invariant tori) lying in between the
chaotic trajectories.} \label{abe:fig11}
\end{center}
\end{figure}

We have found that there is an inherent limitation in the size of the cobweb.
It does not relate to the inevitably finite time of numerical simulations,
which places a practical limit on the
distance over which the
transport can be followed, but is characteristic of the cobweb itself. Our
numerical simulations show (Fig.\ \ref{abe:fig11}) that, for the given parameters, the inner
two-and-a-half loops of the web are distinctly {\it separated} from
the adjacent outer one-and-a-half loops by regular trajectories. This might possibly
be accounted for theoretically by consideration of higher-order approximations
of the averaging method \cite{Bogolyubov:61}. We may speculate that such an
approach could show that, instead of a single infinite cobweb skeleton, there
are many separate separatrices (of the one-and-a-half loop shape) lying closely
together, but that they might then coalesce due to the chaotic layers dressing
them as a result of the perturbation. Because the width of the layer decreases
exponentially fast with increasing distance from the centre, this would mean
that coalescence would occur only within a few inner loops. Just this is observed in
Fig.\ \ref{abe:fig11}, even despite that $\epsilon$ is moderate rather than small.

One may reasonably ask: {\it Is there any subtle way to substantially increase
the size of the web and to enhance transport through it?}

In order to answer this question, let us recall the reason for the exponential
narrowness of the chaotic layer. It follows from Fig.\ \ref{abe:fig8} that it
is attributable to the frequency of the perturbation $\tilde{V}_f$ being much
higher than the eigenfrequency of the unperturbed resonant Hamiltonian
$\tilde{H}_s$. It is clear from Fig.\ \ref{abe:fig8} that the width of the
layer would be much larger if we could manage to modify the original system in
such a way that a new perturbation of the resonance Hamiltonian had a component
whose frequency was of the order of, or less than, the eigenfrequency of the
resonance Hamiltonian $\tilde{H}_s$. In fact, this may readily be accomplished
in at least two different ways: (i) one can {\it add} to the original plane
wave a small perturbation of the slightly shifted frequency (it can itself be
e.g.\ a plane wave); (ii) one can modulate weakly the {\it angle} of the
original plane wave at a low frequency. We demonstrate below only the second
option (it will be especially convenient for realization of the phenomena in
SLs, as shown in Section \ref{superlattices} below).

\begin{figure}[tb]
\begin{center}
\includegraphics*[width = 7.5 cm]{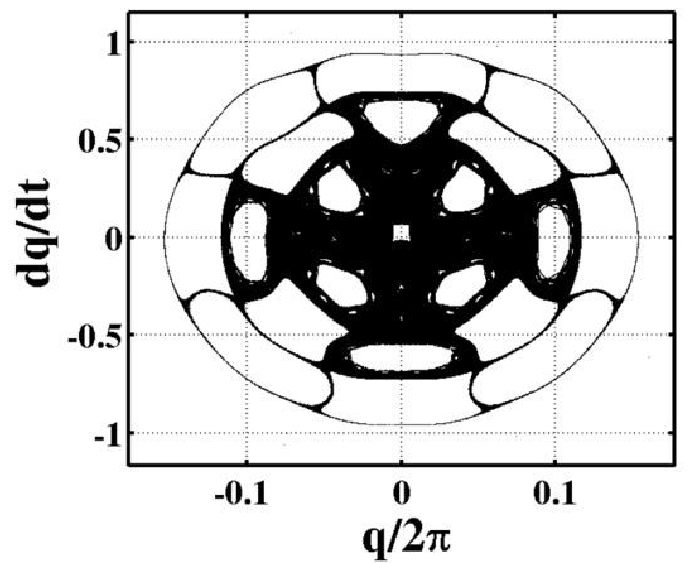} \hspace{0.3cm}
\includegraphics*[width = 7.5 cm]{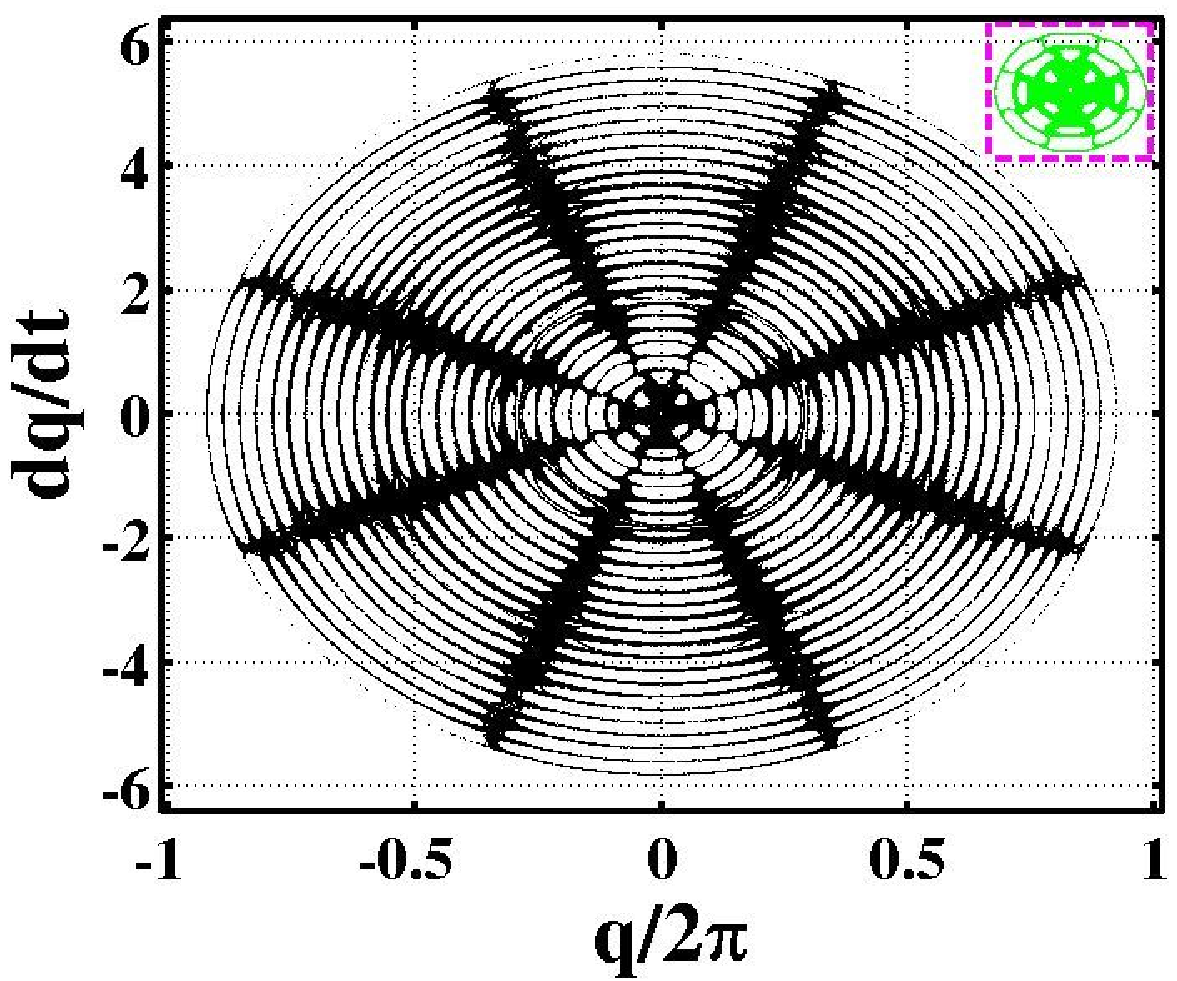}
\caption {Poincar{\'e} section for a trajectory of the system (28) with initial
state $q=0.1,\;\dot{q}=0$ (at instants $t_n=nT$ where $T\equiv 2 \pi/0.02$ is
the period of the modulation and $n=1,2,3,...600000$) for $h=0$ (left panel)
and $h=0.1$ (right panel). A symplectic integration scheme of the fourth order
is used, with an integration step $t_{int}=\frac{2 \pi}{40000}\approx
1.57\times 10^{-4}$, so that the inaccuracy at each step is of the order of
$t_{int}^5\approx \times 10^{-19}$. The left panel corresponds to the
conventional case considered in \cite{Zaslavsky:07,Chernikov:87a,Zaslavsky:91}.
The right panel demonstrates that the modulation, although weak, greatly
enlarges the web size (note the different axes scales), thereby greatly
enhancing the chaotic transport. The inset in the top right hand-corner plots
the left-hand panel on the same scale, thereby illustrating the dramatic extent
of this enlargement.} \label{abc:fig12}
\end{center}\end{figure}

\begin{figure}[tb]
\begin{center}
\includegraphics*[width = 7.3 cm]{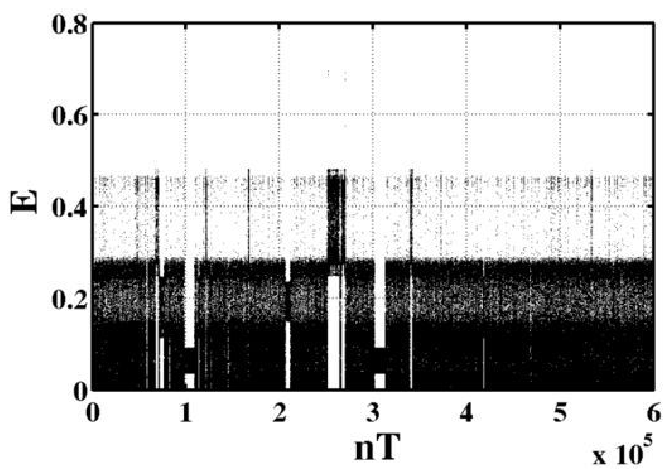} \hspace{0.3cm}
\includegraphics*[width = 7.3 cm]{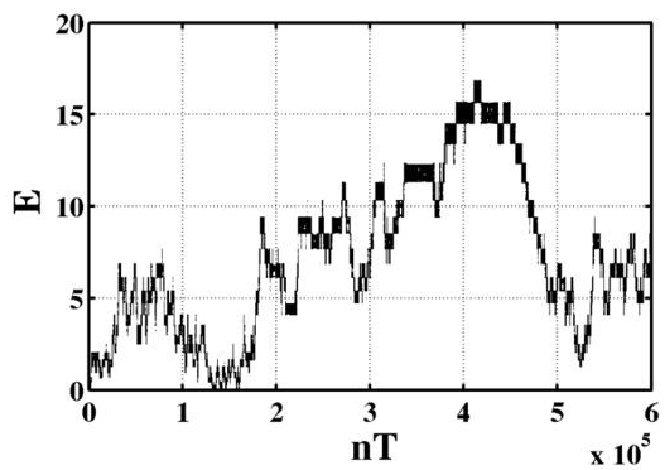}
\caption {Dynamics of the energy for the same systems as in Fig.\
\ref{abc:fig12}.} \label{abc:fig13}
\end{center}
\end{figure}

We therefore consider the modified system (cf.\ the original Eq. (15)):
\begin{eqnarray}
&&
\ddot{q} + \omega_0^2q=\epsilon\frac{\omega_0^2}{k}\sin(kq-\nu t-h\sin(\Omega t)),
\\
&&
\nu=n\omega_0,\quad n=1,2,3, . . .,
\nonumber
\\
&&
h\ll 1,\quad \Omega\stackrel{\sim}{<}\omega_{\rm unperturbed}\sim\frac{\epsilon\omega_0}{I^{3/4}}.
\nonumber
\end{eqnarray}
\noindent Of course, the latter inequality cannot be satisfied for an
arbitrarily large $I$, but it can be true for a sufficiently high value
of $I$ which greatly exceeds the original cobweb size in terms of $I$.

Repeating the same procedure used above in the derivation of Eq.\ (19), i.e.\
transforming to action-angle variables, introducing the slow angle
$\tilde{\theta}$ and the auxiliary Hamiltonian $\tilde{H}\equiv nH-\nu
\tilde{I}$ which governs the dynamics of $\{\tilde{I}-\tilde{\theta}\}$, we can
derive:
\begin{eqnarray}
&&
\tilde{H}=\tilde{H}_s^{\rm (modified)} +\tilde{V}_f,
\\
&&
\tilde{H}_s^{\rm (modified)} \approx\tilde{H}_s +h\frac{\epsilon n}{k^2}\omega_0^2J_n(k\rho(\tilde{I}))\sin(\tilde{\theta})\sin(\Omega t)
\nonumber
\end{eqnarray}
\noindent (in the derivation, we took into account in particular the smallness
of $h$).

Unlike the original autonomous slow resonance Hamiltonian $\tilde{H}_s \equiv
\tilde{H}_s (\tilde{I},\tilde{\theta})$, the modified slow part of the
Hamiltonian, i.e.\ $ \tilde{H}_s^{\rm (modified)} $, depends on time: it contains a
term $\propto h$ which oscillates at a low frequency $\Omega$. It is this
slowly oscillating additional term (rather than the former fast-oscillating
perturbation term $\tilde{V}_f$) that now determines the width of the layer:
the width is moderately small (due to the smallness of $h$ and $\epsilon$),
rather than exponentially small as in the original setup. This exponential
growth in the width of the layer gives rise to substantial growth in the size
of the cobweb.

To illustrate the above ideas, we use the following example:
\begin{equation}
\ddot q + q = 0.1\, \sin[15q -4t -h\sin(0.02\,t)] .
\end{equation}
\noindent For $h=0$, this coincides with the conventional cobweb example
developed in \cite{Zaslavsky:07,Zaslavsky:91,Chernikov:87a}.

Comparison of the left and right panels of Fig.\ \ref{abc:fig12}, corresponding
to $h=0$ and $h=0.1$ respectively, reveals a 6-fold increase in the size of the
web in terms of $q$ and $p\equiv{\rm d}q/{\rm d}t$:
\begin{equation}
n_{q,p}\approx 6.
\end{equation}
We emphasize that the modulation giving rise to this substantial increase is
actually very small: its amplitude of 0.1 is about 60 times smaller than $2\pi$
which is the relevant scale for the angle.

The corresponding increase of the size in terms of energy is proportional to
the square of $n_{q,p}$:
\begin{equation}
n_E=n_{q,p}^2\approx 36.
\end{equation}
\noindent Fig.\ \ref{abc:fig13} shows this explicitly and, in addition,
demonstrates that the mode of transport is significantly changed.

\begin{figure}[tb]
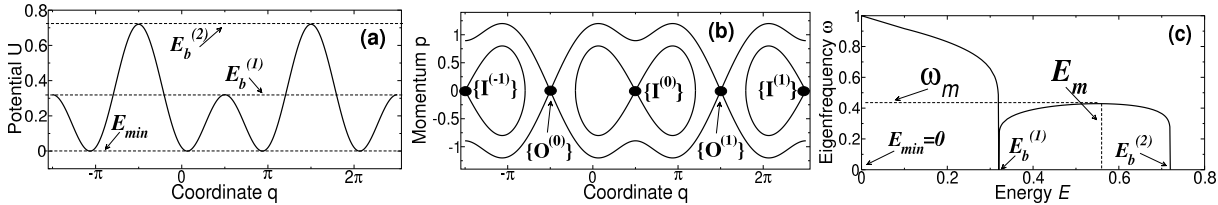

\includegraphics*[width = 5.3 cm]{Fig13a.eps}
\includegraphics*[width = 5.3 cm]{Fig13b.eps}
\includegraphics*[width = 5.3 cm]{Fig13c.eps}
\caption {(a) The potential $U(q)=(0.2-\sin(q))^2/2$, (b) the separatrices in
the phase space, and (c) the frequency of oscillation as a function of energy
$\omega(E)$ for the autonomous potential system $H_0(p,q,)=p^2/2+U(q)$.}
\label{abc:fig14}
\end{figure}

\begin{figure}[tb]
\center{\includegraphics*[width = 10 cm]{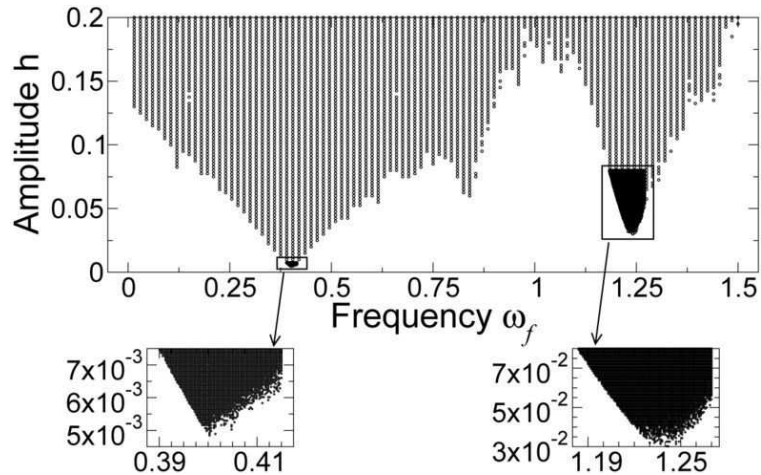}}
\caption {The bifurcation diagram in the plane of amplitude and frequency of
perturbation for the system $H=H_0+hq\cos(\omega_f t)$ with $H_0$ as in Fig.\
\ref{abc:fig14}. The area of $\{h,\omega_f\}$ for which there is a global chaos
between the separatrices is shaded. The lower boundary of the shaded area
therefore corresponds to the onset of global chaos, representing the function
$h_{cr}(\omega_f)$.} \label{abc:fig15}
\end{figure}

\subsection{Inexact resonance}

Our idea of an additional small modulation of the angle of the plane wave is
equally fruitful in the case of an inexact resonance. The frequency band
(around the resonance) in which the web-like structure is formed may grow
exponentially: instead of the exponentially narrow band found in the absence of
modulation, we may have a moderately narrow band.

Moreover, there is a nontrivial spectral dependence of this growth: it reflects
a universal mechanism for facilitation of the onset of chaos between adjacent
separatrices, discovered recently by Soskin, Mannella and Yevtushenko
\cite{Soskin:08}. To explain this mechanism, we use their example:
it is a potential system with a spatially periodic
potential possessing two barriers of different height within one period (Fig.\
\ref{abc:fig14}(a)). Naturally, there are two kinds of separatrices (Fig.
\ref{abc:fig14}(b)). It was shown \cite{Soskin:08} that the frequency of
oscillation $\omega$ as a function of energy $E$ possesses a local maximum
$\omega_m$ between the separatrices and, moreover, $\omega$ is close to $\omega_m$ over most of the
inter-separatrix energy range (Fig.\ \ref{abc:fig14}(c)). The latter property
is valid for any system with two or more separatrices and is particularly
important in the present context.

\begin{figure}[tb]
\center{\includegraphics*[width = 9. cm]{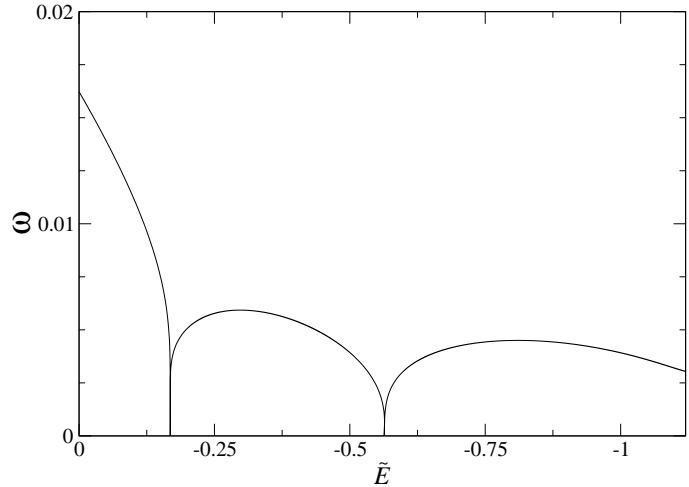}}
\caption {Frequency of oscillation in the autonomous Hamiltonian system (23)
with parameters as in (31), as a function of the corresponding energy
$\tilde{E}\equiv\tilde{H}_s$.} \label{abc:fig16}
\end{figure}

\begin{figure}[tb]
\center{\includegraphics*[width = 9. cm]{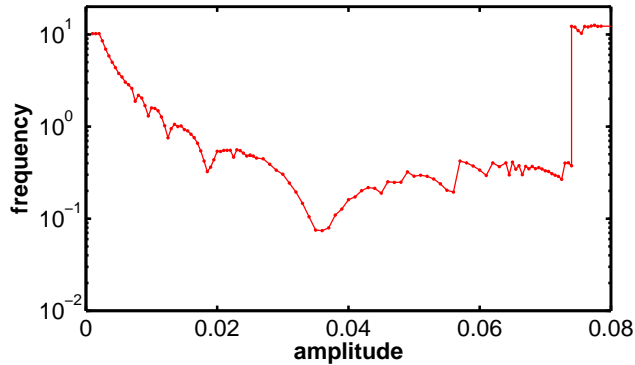}}
\caption {The spectral dependence (i.e.\ dependence on $\Omega$) of the
critical amplitude $h$ of the modulation required for initial web formation
i.e.\ for chaotic connection of the first two separatrices of $\tilde{H}_s$
(23) to occur. Note the logarithmic vertical scale.} \label{abc:fig17}
\end{figure}

If we perturb the system with a time-periodic perturbation of frequency
slightly lower than $\omega_m$ then, due to the flatness of $\omega(E)$ over
most of the inter-separatrix range of $E$, two nonlinear resonances arise that
are very wide in terms of energy. Even a rather small amplitude of perturbation
may be sufficient for these nonlinear resonances to overlap with
each other and with the separatrix chaotic layers, thus connecting the
latter by the chaotic transport. This has been confirmed both theoretically and
in numerical simulations. Consequently, a critical perturbation amplitude
$h_{cr}$ is required for chaotic transport between the separatrix chaotic layers
(which may
be considered as the onset of global chaos between them). As a
function of the perturbation frequency $\omega_f$, it possesses a deep
minimum\footnote{If the perturbation is parametric rather than additive, then
the deepest minimum may occur at some multiple of $\omega_m$  rather than at
$\omega_m$ itself \cite{Soskin:08}.} at a frequency approximately equal to
$\omega_m$. This is not only seen in the simulations (Fig.\ \ref{abc:fig15})
but is also well described by the theory \cite{Soskin:08}.

The situation is similar for modulation-assisted formation of the web in the
case of inexact resonance between the plane wave frequency and that of the
oscillator. To demonstrate this, we use the following example (the parameters
correspond to those used in experiments on semiconductor SLs):
\begin{eqnarray}
&&\ddot q + q = \epsilon \sin[q -\nu t -h\sin(\Omega t)] ,
\\
&&
\nu=1.02292, \quad\epsilon=0.573.
\nonumber
\end{eqnarray}
\noindent For $h=0$, a stochastic web is not formed because the $\Delta
\omega\equiv \nu-1\approx 0.023$ is too large for the chaotic connection of any
separatrices of $\tilde{H}_s$ (23) to occur. We have calculated numerically the
two lowest separatrices in the plane $\tilde{I}-\tilde{\theta}$ (cf.\ Fig.\
\ref{abe:fig9}), and then obtained the frequency $\omega$ of oscillation of
$\tilde{I}$ (or, equivalently, of the shift by $2\pi$ of $\tilde{\theta}$) as a
function of the auxiliary energy $\tilde{E}\equiv\tilde{H}_s$: see Fig.\
\ref{abc:fig16}. There is a local maximum that is clearly similar to that in
Fig.\ \ref{abc:fig14}(c).

Then we switch on the modulation of the wave angle and, for each given
$\Omega$, increase $h$ gradually, until the web is formed i.e.\ until chaotic
connection occurs between the first two separatrices $\tilde{H}_s$. This may be
considered as the formation of the web. The spectral dependence of the
corresponding critical amplitude is shown in Fig.\ \ref{abc:fig17}. Similar to
Fig.\ \ref{abc:fig15}, it exhibits a deep minimum (note the logarithmic scale)
at a frequency which is a little smaller than the local maximum of the
dependence $\omega(\tilde{E})$: cf. Fig. 16.

\section {Semiconductor superlattices in electric and magnetic
fields}\label{superlattices}

\begin{figure}[tb]
\center{\includegraphics*[width = 8 cm]{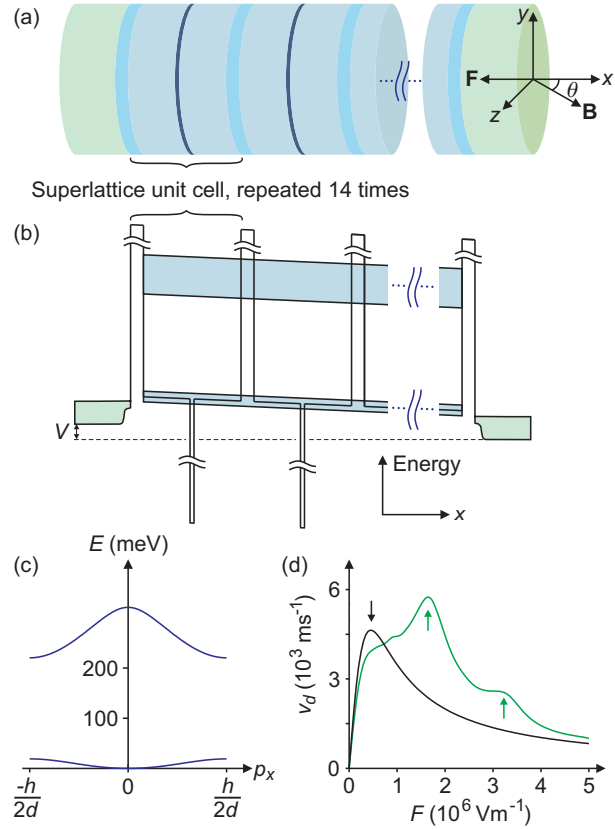}}
\caption {(a). Schematic diagram of the SL. Its unit cell comprises two
3.5-nm-thick Ga-As layers (light blue), a 0.3-nm-thick InAs layer (dark blue),
and a 1-nm-thick AlAs barrier layer (mid blue). The structure contains 14 unit
cells, enclosed by 50-nm-thick GaAs ohmic contacts (green). (b). Schematic
variation of the electronic potential energy with position $x$ normal to the
layers, for $V\neq 0$. The quantum wells produce a periodic potential (only two
complete unit cells are shown for clarity supporting two minibands (blue)).
Green areas represent electron gases in the contacts. (c). Energy versus
crystal momentum dispersion curves for the two minibands. (d). Plots of the
drift velocity $v_d$ versus $F$ calculated for $B=11{\rm T}$ with $\theta=0$
(black curve: arrow marks peak) and $\theta=45^{\circ}$ (green curve: arrows mark
additional peaks). Reprinted by permission from Macmillan Publishers Ltd: [Nature] \cite{Fromhold:04}, copyright (2004).} \label{abc:fig18}
\end{figure}

An application of the stochastic cobweb in nanoscience was recently identified
and discussed in a series of publications by researchers from the University of
Nottingham \cite{Fromhold:01,Fromhold:04,Fowler:07,Balanov:08,Greenaway:09}.
They considered quantum electron transport in nanometre-scale 1D semiconductor
SLs subject to a constant electric field along the SL axis and to a constant
magnetic field (Fig.\ \ref{abc:fig18}(a,b)). The spatial periodicity of the SL
layers gives rise to minibands for the electrons (Fig.\ \ref{abc:fig18}(c)). In
the tight-binding approximation, the electron's energy $E$ as a function of its
momentum $\vec{p}$ in the lowest miniband is given by
\cite{Fromhold:01,Balanov:08}
\begin{equation}
E(\vec{p})=\frac{\Delta[1-\cos(p_xd/\hbar)]}{2}+\frac{p_y^2+p_z^2}{2m^{*}},
\end{equation}
\noindent where $x$ is oriented along the SL axis, $\Delta$ is the miniband
width, $d$ is the SL period, and $m^{*}$ is the electron effective mass for
motion in the transverse (i.e.\ $y$-$z$) direction.

Thus, the quasi-classical motion of an electron of charge $e$ in an electric
field $\vec{F}$ and a magnetic field $\vec{B}$ can be described by:
\begin{equation}
\frac{{\rm d}\vec{p}}{{\rm d}t}=-e\{\vec{F} + [\nabla_{\vec{p}}
E(\vec{p}) \times \vec{B}]\}.
\end{equation}
It was shown in \cite{Fromhold:01} that, for a constant electric field along
the SL axis $\vec{F}=(-F,0,0)$ and constant magnetic field with a given angle
$\theta$ to the axis $\vec{B}=(B\cos(\theta),0,B\sin(\theta))$, the dynamics of
the $z$-component of momentum $p_z$ reduces to the equation of motion of an
auxiliary harmonic oscillator subject to a plane wave i.e.\ to the equation
considered in the previous sections\footnote{The only small difference is the
presence of a constant shift $\phi $ in the wave angle, but it is
inessential.}$^,$\footnote{The motion of electrons in a biased SL with a tilted
magnetic field can also be linked to the ultra-fast Fiske effect observed for a
Josephson junction coupled to an electromagnetic resonator
\cite{Kosevich:06}.}:
\begin{eqnarray}
&&
\ddot{p_z} + \omega_0^2p_z=\epsilon\frac{\omega_0^2}{k}\sin(kp_z-\nu t+\phi),
\\
&&
\omega_0=\omega_c\cos(\theta),\quad \omega_c\equiv\frac{Be}{ m^{*}},
\nonumber
\\
&&
\nu=\omega_B, \quad\omega_B\equiv\frac{eFd}{\hbar},
\nonumber
\\
&&
\epsilon=\frac{\Delta m^{*}d^2\tan^2(\theta)}{ \hbar^2},\quad
\nonumber
\\
&&
k=\frac{ d \tan(\theta)}{ \hbar} ,
\nonumber \\
&&
\phi=\pi+\frac{ d}{ \hbar}[p_x(t=0)+p_z(t=0) \tan(\theta)].
\nonumber
\end{eqnarray}
\noindent We emphasize that, despite its classical appearance, Eq.\ (34) has an
inherently quantum origin: most of the parameters contain Planck's constant
$\hbar$.

\begin{figure}[tb]
\center{\includegraphics*[width = 7 cm]{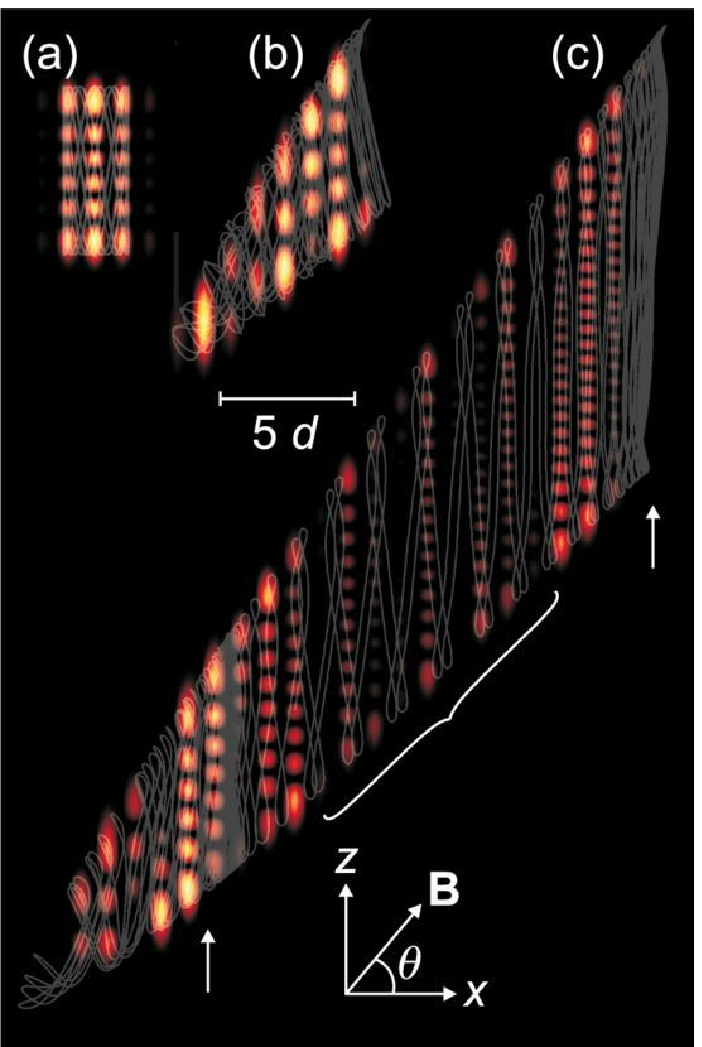}}
\caption {Electron trajectories and wavefunctions. Grey translucent curves:
classical electron trajectories in the $x$-$z$ plane (axes inset) overlaid on
corresponding plots of $|\Psi(x,z)|^2$ (black zero, yellow high) at $B=11{\rm
T}$. (a). When $\theta=0$, the probability distribution is concentrated within
the turning points of the classical trajectory. (b). Off resonance, here for
$r=(1+\sqrt{5})/4)$ and $\theta=50^{\circ}$, the trajectories and wave functions
extend a little but are still quite localized. (c). On resonance, here for
$r=1$ and $\theta=50^{\circ}$, the wave functions extend across many SL periods, in
correspondence with the extended classical trajectories. A region of high
probability density (yellow peaks) associated with a concentration of orbital
loops, occurs when the electron is trapped on the first (inner) ring-shaped
filament of the web (lower left-hand arrow marks the $x$ value corresponding to
this ring) and is therefore unable to progress through the SL. But when the
electron transfers onto the quasi-linear filaments, it shifts rapidly along $x$, following widely spaced orbital loops (within bracket), which correspond to
low probability density. The wavefunction is bounded from the right by the
second ring-shaped web filament (right-hand arrow marks $x$ position
corresponding to this ring), which impedes electron flow. Reprinted by permission from Macmillan Publishers Ltd: [Nature] \cite{Fromhold:04}, copyright (2004).} \label{abc:fig19}
\end{figure}

The dynamics of the system is fully determined by the dynamics of $p_z$. Fig.\
\ref{abc:fig19} shows how the trajectory of an electron in the $x$-$z$ plane
changes with the angle of the magnetic field.

At $\theta=0$, the plane wave has zero amplitude and the motion along the $x$-
and $z$-directions is separable. Electrons undergo Bloch oscillations along $x$
(due to the presence of the constant electric field) and cyclotron motion about
$\vec B$ (Fig.\ \ref{abc:fig19}(a)). The motion is localized.

Tilting $\vec B$ produces nonlinear coupling of the Bloch and cyclotron motion:
as $\theta\neq 0$, the plane wave in (34) acquires a non-zero amplitude. This
causes some moderate delocalization of trajectories (Fig.\ \ref{abc:fig19}(b)).
The delocalization grows very fast (Fig.\ \ref{abc:fig19}(c)) when $\theta$
reaches values corresponding to the integer values $r\equiv
\omega_B/(\omega_c\cos(\theta))$, in other words to the resonance
$\nu=n\omega_0$. This strong delocalization in $x$ is a consequence of the
onset of the stochastic web for the motion of $p_z$ (34).

It is remarkable that the quantum probability density $|\Psi(x,z)|^2$,
calculated by solution of the Schr\"{o}dinger equation in the SL model potential
subjected to electric and magnetic fields should so nicely follow
quasi-classical trajectories based on the dynamics of $p_z$ (34): see Fig.\
\ref{abc:fig19}.

\begin{figure}[tb]
\center{\includegraphics*[width = 7 cm]{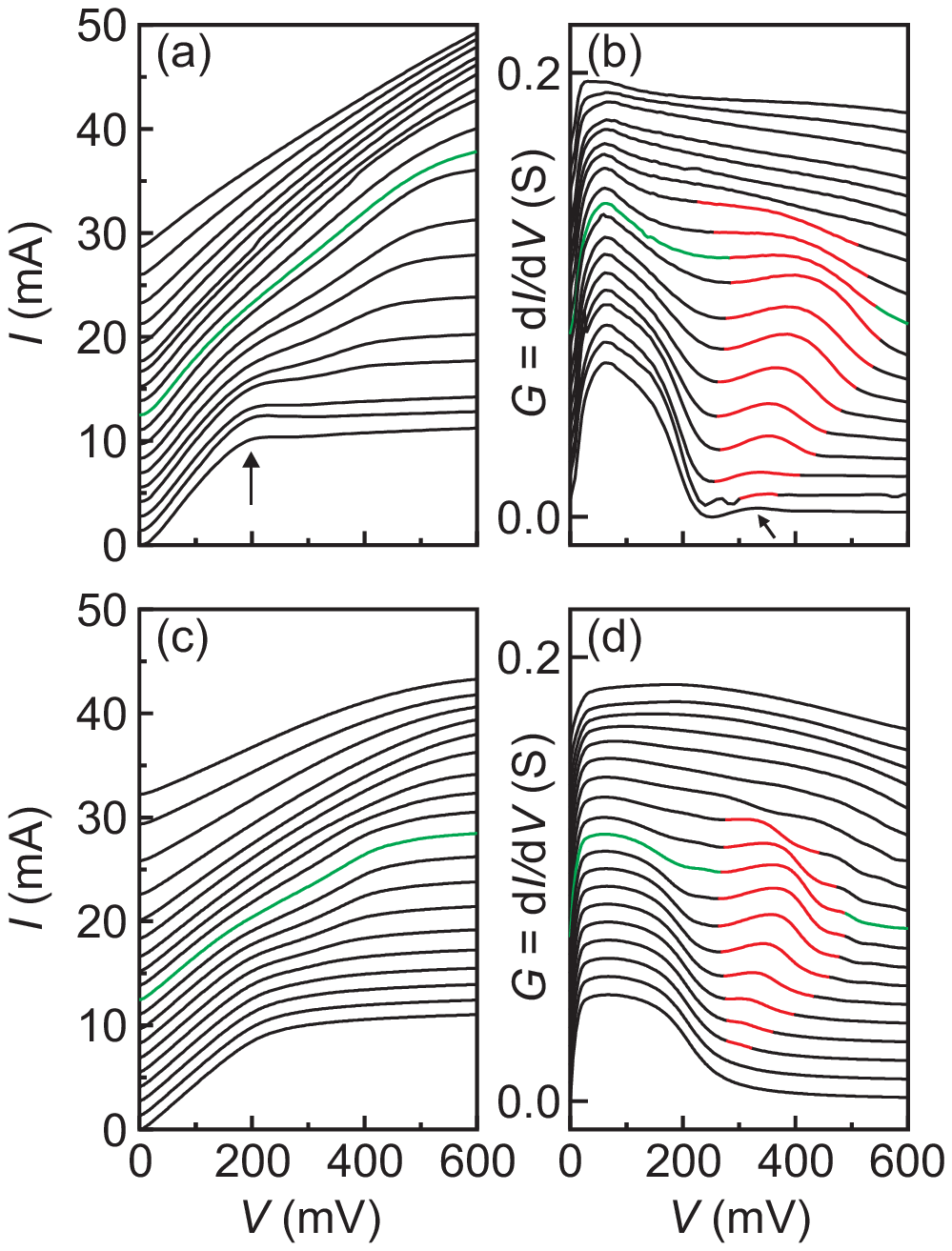}}
\caption {Resonant enhancement of current. (a). Experimental $I(V)$ curves
measured for $B=11{\rm T}$ and $\theta=0$ (bottom curve) to $90^{\circ}$ (top curve)
at $5^{\circ}$ intervals at a lattice temperature of 4.2K. For
$10^{\circ}\leq\theta\leq55^{\circ}$, each curve contains a region of enhanced $I$ beyond
$V\approx 250$mV. (b) Differential conductance plots of the data in (a) reveal
strong resonant peaks (red). (c). Theoretical $I(V)$ characteristics (for same
parameters as in (a)). (d). Differential conductance plots of the traces in
(c). Curves in (a)-(d) are offset vertically for clarity and those for
$\theta=45^{\circ}$ are green. In theory and experiment, the resonant peaks in $G(V)$
initially shift slightly to higher $V$ as $\theta$ increases, because the
enhanced conductance leads to higher electron charge density in the SL, which
increases $F$ and $V$. Reprinted by permission from Macmillan Publishers Ltd: [Nature] \cite{Fromhold:04}, copyright (2004).} \label{abc:fig20}
\end{figure}

As shown in \cite{Fromhold:01,Fromhold:04}, the delocalization of the
electrons\footnote{Seemingly paradoxically, one must take account of scattering
in the calculation: if the motion were purely Hamiltonian, the position of the
electron averaged over time would be constant.} strongly affects their drift
velocity $v_d$ and, as a consequence, the current $I$ and the current-voltage
dependence $I(V)$. There are clear manifestations of the resonances, both in
the theoretical curves $v_d(F)$, $I(V)$, ${\rm d}I/{\rm d}V$ (see Fig.\
\ref{abc:fig18}(d), Fig.\ \ref{abc:fig20}(c) and Fig.\ \ref{abc:fig20}(d)
respectively), and in the experimental curves $I(V)$ and ${\rm d}I/{\rm d}V$
(Figs.\ \ref{abc:fig20}(a) and (b) respectively). Thus there is clear evidence
for stochastic web formation in quantum electron transport, providing the basis
for a conceptually new method for its control. When scattering is included
\emph{a priori} in the semiclassical equations of motion, the stochastic web,
and stable islands that it enmeshes, evolve into limit cycles. These limit
cycles also exhibit sharp resonant delocalization and their locations in phase
space closely reflect the underlying web topology \cite{Balanov:08}.

Finally, we pose a question: could the modification of the stochastic web
discussed in the previous section be of use for the SLs? Its seems \cite
{Soskin:09b,Soskin:09c,Soskin:09d} that this is indeed the case. As shown in
Section \ref{modstochwebs} above, modulation of the wave angle results in a
large increase of the web size. It was noted in \cite{Fromhold:01} that the
delocalization in $x$ is proportional to the web size in terms of the energy of
the oscillator in $p_z$, i.e.\ to $E=\dot{p_z}^2/2+\omega_0^2p_z^2/2$ (cf.\
Fig.\ \ref{abc:fig19}). The only question is how the SL should be perturbed in
order for the modulation term to appear in the dynamical equation for $p_z$.
One suggestion \cite{Soskin:09b,Soskin:09c,Soskin:09d} is that, in a manner
similar to the derivation of Eq.\ (34), one can show that the modulation term in
the equation for $p_z$ appears if an ac-component is added to the constant
electric field:

\begin {equation}
F\rightarrow F+F_{ac}\cos(\omega_{ac}t).
\end {equation}
Then, the following modulation term is added in the wave angle in Eq.\ (34):
\begin {equation}
h\sin(\omega_{ac}t), \quad\quad h=\frac{F_{ac}}{F}\frac{\omega_B}{\omega_{ac}}.
\end {equation}
To compare the resulting equation with the example (28) that we studied
numerically, we transform to normalized time
\begin {equation}
t\rightarrow \tilde{t}\equiv\omega_0t.
\end {equation}
\noindent Then the equation of motion for $p_z$ is:
\begin{eqnarray}
&&
\frac{{\rm d}^2p_z}{{\rm d}\tilde{t}^2} + p_z= \frac{\epsilon }{k}\sin(kp_z-\nu \tilde{t}+\phi+h\sin(\Omega\tilde{t})),
\\
&&
\Omega=\frac{\omega_{ac}}{\omega_0}, \quad\quad
h=\frac{F_{ac}}{F}\frac{\nu/\omega_0}{\Omega},
\nonumber
\end{eqnarray}
\noindent where all other parameters are as in Eq. (34).

Thus, if the parameters are similar to those in (28), in particular: $h=0.1$,
$\Omega=0.02$, $\nu/\omega_0=4$, then we will have an enlargement in $E$ as
found for Eq.\ (28): $n_E\approx 36$. In order to achieve this, we need
$F_{ac}/F=h\Omega/(\nu/\omega_0)=1/2000$. This means that in order to achieve
delocalization of the electron by a factor of about 40, we need to add to the
constant electric field an ac-component of amplitude that is smaller than the
constant component by a factor of 2000! We remind the reader that the reason
for such a dramatic change when an ac-component is added is the exponentially
strong enhancement of chaotic transport through the stochastic web due to the
modulation of the wave angle.

Recently, the effects of stochastic web formation on the high-frequency
(GHz-THz) performance of the SLs has been considered
\cite{Greenaway:09,Hyart:09}. Modulation of the $v_d(F)$ curves, induced by
stochastic web formation, leads to the formation of multiple propagating
electron accumulation and depletion regions (charge domains), which greatly
increase both the strength and frequency of the associated temporal current
oscillations. Chaos-assisted motion through stochastic webs may, therefore,
provide a mechanism for controlling the collective dynamics of electrons in SLs
and, hence, for enhancing their THz performance by using \emph{single-particle}
miniband transport to tailor the shape of the $v_d(F)$ curves.

\section {Conclusions}\label{conclusions}

We have shown that, in general, there is a possibility for energy in a
Hamiltonian system to be increased from small to rather large values as a
result of transport through a stochastic web.

In a multi-dimensional system, the onset of a stochastic web is a common
phenomenon, predicted by Arnold in 1964. In the present review, we have been
more interested in the low-dimensional stochastic webs discovered by Chernikov
et al.\ in the late 1980s. They occur in special situations: in a harmonic,
or nearly harmonic, oscillator driven by perturbations periodic in time and
space that are resonant, or nearly resonant, with the oscillator.

We emphasized that the stochastic cobweb can arise when the oscillator is
driven by a weak resonant, or nearly resonant, plane wave. The exponentially
small width of a strand of the web is a characteristic feature of all
stochastic webs and it decreases exponentially fast as the distance from the
centre of the cobweb increases. Moreover there is an inherent limitation on the
size of the cobweb. Soskin et al.\ have suggested how to overcome the
restriction in size of the cobweb and the exponential narrowness of its chaotic
layer, just by slightly modifying the system by means of a small slow
modulation of the angle of the plane wave.

The model of the stochastic web turned out to be directly relevant to the
quantum transport of electrons in semiconductor SLs in constant electric and
magnetic fields, as demonstrated by Fromhold et al.: the quantum transport
dynamics reduces to the model of the harmonic oscillator perturbed by a plane
wave, where parameters are determined by the values of the electric and
magnetic fields, by the angle between them, by the period of the SL, by the
charge and the effective mass of the electron, and by Planck's constant. At
certain values of the parameters, in particular of the electric field,
resonance occurs between the oscillator and the plane wave, resulting in the
onset of the stochastic cobweb and, consequently, in a strong delocalization of
the electron which, in turn, increases the current and gives rise to a peak in
the dependence of the differential conductivity on voltage.

An addition to the constant electric field of a small slow ac-component results
in the slow modulation of the plane angle and, therefore, promises to strongly
increase the delocalization of the electron and to enhance a range of related
phenomena.

\section*{Acknowledgements}

The authors gratefully acknowledge financial support from the Royal Society of
London, the International Centre for Theoretical Physics (Trieste), Pisa
University and the Engineering and Physical Sciences Research Council (UK).
PVEMcC acknowledges the hospitality of the Institute of Semiconductor Physics
during his visit to Kiev, during which the review was conceived. SMS
acknowledges the hospitality of Lancaster University, where work relevant to
some parts of the review was carried out, the hospitality of the University of
Nottingham, where he discussed relevant issues during his visit, and the
hospitality of Pisa University, where the first draft of the review was
prepared.

\vspace{2cm}

\section*{Notes on contributors}

\begin{minipage}[b]{.7\linewidth}Stanislav M.\ Soskin graduated from Kiev State University
in 1982 and obtained PhD from the Institute of Semiconductor Physics (Kiev,
Ukraine) in 1988. At present, he is a Leading Scientific Researcher in the
Theoretical Physics Department of the Institute of Semiconductor Physics. He is
also an Associate Member of the International Centre for Theoretical Physics
(Trieste, Italy) and a visiting member of the Physics Department in Lancaster
University. He is an author of about 80 papers, including a number of reviews,
mainly in the areas of fluctuation phenomena and nonlinear dynamics.
\end{minipage}
\hspace*{.71\linewidth}\begin{minipage}{.29\linewidth}
\vspace*{-4.8cm}\includegraphics[width=\linewidth]{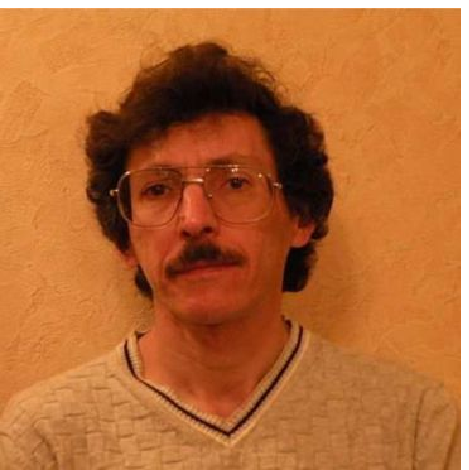}
\end{minipage}

\vspace{0.3cm}

\noindent \begin{minipage}[b]{.7\linewidth}Peter V.E.\ McClintock is Professor
of Physics at Lancaster University. After his education at Queen's University
Belfast and Oxford University, followed by postdoctoral research at Duke
University, he came to Lancaster in 1968. He was a SERC/EPSRC Senior Fellow,
1990-1995. His research interests include low temperature physics,
superfluidity, quantum turbulence, nonlinear dynamics and, most recently, the
applications of nonlinear dynamics to biomedical problems.
\end{minipage}
\hspace*{.71\linewidth}\begin{minipage}{.29\linewidth}
\vspace*{-3.5cm}\includegraphics[width=\linewidth]{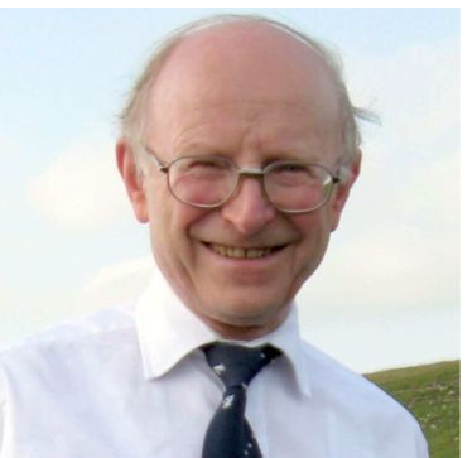}
\end{minipage}

\vspace{0.3cm}
\newpage
\noindent \begin{minipage}[b]{.7\linewidth}T.\ Mark Fromhold was born in 1965
in York, England. He was awarded a first-class honours degree in physics from
the University of Durham (1986) and a PhD in condensed matter theory from the
University of Nottingham (1990). He worked as a post-doctoral research
assistant at the University of Warwick and as a medical physicist at Lincoln
County Hospital (1991). He was a visiting scientist at the National Research
Council (NRC) Ottawa (1995/1996) and held a Gordon Godfrey Fellowship at the
University of New South Wales (1996). In 1995, he was awarded a 5 year EPSRC
Advanced Fellowship. At the end of this Fellowship, he became a Lecturer in
Physics at the University of Nottingham and was promoted to Reader in
Theoretical Physics (2001) and Professor of Physics (2004). His research
interests include quantum transport and chaos in semiconductor
heterostructures, ultracold atoms, and atom chips.
\end{minipage}
\hspace*{.71\linewidth}\begin{minipage}{.29\linewidth}
\vspace*{-8.5cm}\includegraphics[width=\linewidth]{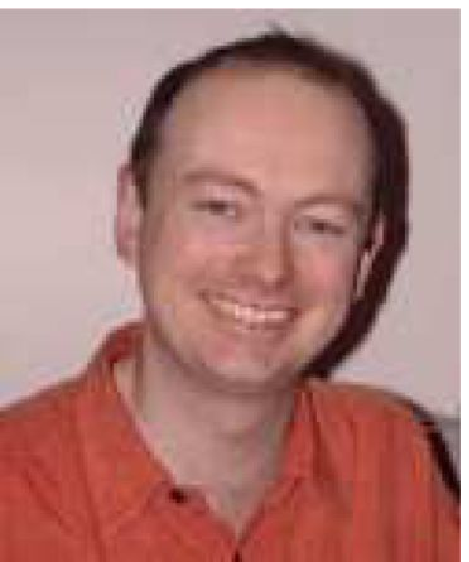}
\end{minipage}

\vspace{0.3cm}

\noindent \begin{minipage}[b]{.7\linewidth}Igor A.\ Khovanov is an EPSRC
Advanced Fellow and Lecturer in the School of Engineering in the University of
Warwick. Following his undergraduate education, PhD, and academic position at
Saratov State University he was a Humboldt Research Fellow in the Humboldt
University in Berlin, and then EPSRC Advanced Fellow in Lancaster before moving
to his permanent position in Warwick. His research interests include nonlinear
dynamics and fluctuation theory and their applications to engineering and
biomedical problems ranging from the stress dynamics of dilute alloys to the
modelling of biological ion channels.
\end{minipage}
\hspace*{.71\linewidth}\begin{minipage}{.29\linewidth}
\vspace*{-4.6cm}\includegraphics[width=\linewidth]{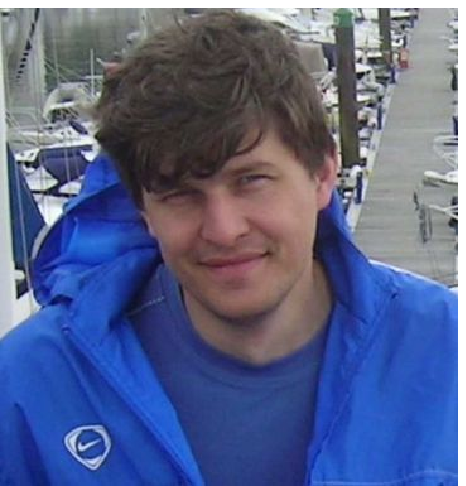}
\end{minipage}

\vspace{0.4cm}

\noindent \begin{minipage}[b]{.7\linewidth}Riccardo Mannella was born in Italy
in 1960. After completing his Laurea degree in Pisa, and a PhD in Lancaster, he
was a postdoctoral research associate both in Lancaster and in Pisa. From 1992
he was a Physics Researcher in Pisa University, and in 2005 he was appointed to
his present position as Professor of Physics in the Veterinary Faculty of Pisa
University. His research interests centre on nonlinear dynamics and fluctuation
theory and their application to a wide variety of problems, including tunneling
of Bose-Einstein condensates in optical lattices, the nonlinear Wannier-Stark
problem and the stock market.
\end{minipage}
\hspace*{.71\linewidth}\begin{minipage}{.29\linewidth}
\vspace*{-4.7cm}\includegraphics[width=\linewidth]{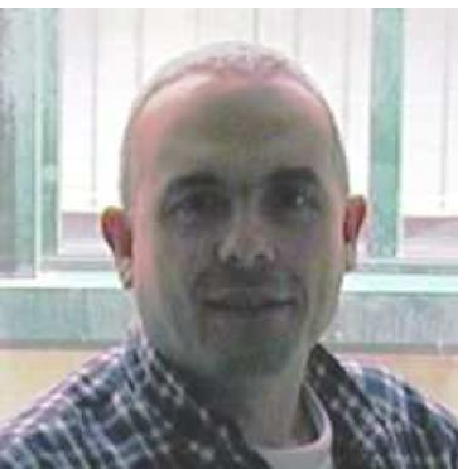}
\end{minipage}


\end{document}